\documentclass[11pt]{article}

    \usepackage[T1]{fontenc}
    \usepackage{mathpazo}
\usepackage{pbox}

    \usepackage{graphicx}
    \makeatletter
    \def\maxwidth{\ifdim\Gin@nat@width>\linewidth\linewidth
    \else\Gin@nat@width\fi}
    \makeatother
    \let\Oldincludegraphics\includegraphics
    \renewcommand{\includegraphics}[1]{\Oldincludegraphics[width=.8\maxwidth]{#1}}
    \usepackage{caption}
    \DeclareCaptionLabelFormat{nolabel}{}
    \usepackage{adjustbox} 
    \usepackage{xcolor} 
    \usepackage{enumerate} 
    \usepackage{geometry} 
    \usepackage{amsmath} 
    \usepackage{changepage}
    \usepackage{amssymb} 
    \usepackage{textcomp} 
    \AtBeginDocument{%
    }
    \usepackage{upquote} 
    \usepackage{eurosym} 
    \usepackage[mathletters]{ucs} 
    \usepackage[utf8x]{inputenc} 
    \usepackage{fancyvrb} 
    \usepackage{grffile} 
    \usepackage{hyperref}
    \usepackage{longtable} 
    \usepackage{booktabs}  
    \usepackage[inline]{enumitem} 
    \usepackage[normalem]{ulem} 
    \usepackage{geometry}

    \definecolor{urlcolor}{rgb}{0,.145,.698}
    \definecolor{linkcolor}{rgb}{.71,0.21,0.01}
    \definecolor{citecolor}{rgb}{.12,.54,.11}

    \definecolor{ansi-black}{HTML}{3E424D}
    \definecolor{ansi-black-intense}{HTML}{282C36}
    \definecolor{ansi-red}{HTML}{E75C58}
    \definecolor{ansi-red-intense}{HTML}{B22B31}
    \definecolor{ansi-green}{HTML}{00A250}
    \definecolor{ansi-green-intense}{HTML}{007427}
    \definecolor{ansi-yellow}{HTML}{DDB62B}
    \definecolor{ansi-yellow-intense}{HTML}{B27D12}
    \definecolor{ansi-blue}{HTML}{208FFB}
    \definecolor{ansi-blue-intense}{HTML}{0065CA}
    \definecolor{ansi-magenta}{HTML}{D160C4}
    \definecolor{ansi-magenta-intense}{HTML}{A03196}
    \definecolor{ansi-cyan}{HTML}{60C6C8}
    \definecolor{ansi-cyan-intense}{HTML}{258F8F}
    \definecolor{ansi-white}{HTML}{C5C1B4}
    \definecolor{ansi-white-intense}{HTML}{A1A6B2}

    \providecommand{\tightlist}{%
      \setlength{\itemsep}{0pt}\setlength{\parskip}{0pt}}
    \DefineVerbatimEnvironment{Highlighting}{Verbatim}{commandchars=\\\{\}}

\makeatletter
\def\PY@reset{\let\PY@it=\relax \let\PY@bf=\relax%
    \let\PY@ul=\relax \let\PY@tc=\relax%
    \let\PY@bc=\relax \let\PY@ff=\relax}
\def\PY@tok#1{\csname PY@tok@#1\endcsname}
\def\PY@toks#1+{\ifx\relax#1\empty\else%
    \PY@tok{#1}\expandafter\PY@toks\fi}
\def\PY@do#1{\PY@bc{\PY@tc{\PY@ul{%
    \PY@it{\PY@bf{\PY@ff{#1}}}}}}}
\def\PY#1#2{\PY@reset\PY@toks#1+\relax+\PY@do{#2}}

\expandafter\def\csname PY@tok@sc\endcsname{\def\PY@tc##1{\textcolor[rgb]{0.73,0.13,0.13}{##1}}}
\expandafter\def\csname PY@tok@vg\endcsname{\def\PY@tc##1{\textcolor[rgb]{0.10,0.09,0.49}{##1}}}
\expandafter\def\csname PY@tok@cp\endcsname{\def\PY@tc##1{\textcolor[rgb]{0.74,0.48,0.00}{##1}}}
\expandafter\def\csname PY@tok@nf\endcsname{\def\PY@tc##1{\textcolor[rgb]{0.00,0.00,1.00}{##1}}}
\expandafter\def\csname PY@tok@c\endcsname{\let\PY@it=\textit\def\PY@tc##1{\textcolor[rgb]{0.25,0.50,0.50}{##1}}}
\expandafter\def\csname PY@tok@gt\endcsname{\def\PY@tc##1{\textcolor[rgb]{0.00,0.27,0.87}{##1}}}
\expandafter\def\csname PY@tok@kc\endcsname{\let\PY@bf=\textbf\def\PY@tc##1{\textcolor[rgb]{0.00,0.50,0.00}{##1}}}
\expandafter\def\csname PY@tok@sd\endcsname{\let\PY@it=\textit\def\PY@tc##1{\textcolor[rgb]{0.73,0.13,0.13}{##1}}}
\expandafter\def\csname PY@tok@bp\endcsname{\def\PY@tc##1{\textcolor[rgb]{0.00,0.50,0.00}{##1}}}
\expandafter\def\csname PY@tok@err\endcsname{\def\PY@bc##1{\setlength{\fboxsep}{0pt}\fcolorbox[rgb]{1.00,0.00,0.00}{1,1,1}{\strut ##1}}}
\expandafter\def\csname PY@tok@mh\endcsname{\def\PY@tc##1{\textcolor[rgb]{0.40,0.40,0.40}{##1}}}
\expandafter\def\csname PY@tok@gr\endcsname{\def\PY@tc##1{\textcolor[rgb]{1.00,0.00,0.00}{##1}}}
\expandafter\def\csname PY@tok@ni\endcsname{\let\PY@bf=\textbf\def\PY@tc##1{\textcolor[rgb]{0.60,0.60,0.60}{##1}}}
\expandafter\def\csname PY@tok@cs\endcsname{\let\PY@it=\textit\def\PY@tc##1{\textcolor[rgb]{0.25,0.50,0.50}{##1}}}
\expandafter\def\csname PY@tok@mi\endcsname{\def\PY@tc##1{\textcolor[rgb]{0.40,0.40,0.40}{##1}}}
\expandafter\def\csname PY@tok@no\endcsname{\def\PY@tc##1{\textcolor[rgb]{0.53,0.00,0.00}{##1}}}
\expandafter\def\csname PY@tok@fm\endcsname{\def\PY@tc##1{\textcolor[rgb]{0.00,0.00,1.00}{##1}}}
\expandafter\def\csname PY@tok@dl\endcsname{\def\PY@tc##1{\textcolor[rgb]{0.73,0.13,0.13}{##1}}}
\expandafter\def\csname PY@tok@ge\endcsname{\let\PY@it=\textit}
\expandafter\def\csname PY@tok@gi\endcsname{\def\PY@tc##1{\textcolor[rgb]{0.00,0.63,0.00}{##1}}}
\expandafter\def\csname PY@tok@nt\endcsname{\let\PY@bf=\textbf\def\PY@tc##1{\textcolor[rgb]{0.00,0.50,0.00}{##1}}}
\expandafter\def\csname PY@tok@sa\endcsname{\def\PY@tc##1{\textcolor[rgb]{0.73,0.13,0.13}{##1}}}
\expandafter\def\csname PY@tok@sb\endcsname{\def\PY@tc##1{\textcolor[rgb]{0.73,0.13,0.13}{##1}}}
\expandafter\def\csname PY@tok@vc\endcsname{\def\PY@tc##1{\textcolor[rgb]{0.10,0.09,0.49}{##1}}}
\expandafter\def\csname PY@tok@o\endcsname{\def\PY@tc##1{\textcolor[rgb]{0.40,0.40,0.40}{##1}}}
\expandafter\def\csname PY@tok@ow\endcsname{\let\PY@bf=\textbf\def\PY@tc##1{\textcolor[rgb]{0.67,0.13,1.00}{##1}}}
\expandafter\def\csname PY@tok@gd\endcsname{\def\PY@tc##1{\textcolor[rgb]{0.63,0.00,0.00}{##1}}}
\expandafter\def\csname PY@tok@kn\endcsname{\let\PY@bf=\textbf\def\PY@tc##1{\textcolor[rgb]{0.00,0.50,0.00}{##1}}}
\expandafter\def\csname PY@tok@nl\endcsname{\def\PY@tc##1{\textcolor[rgb]{0.63,0.63,0.00}{##1}}}
\expandafter\def\csname PY@tok@nc\endcsname{\let\PY@bf=\textbf\def\PY@tc##1{\textcolor[rgb]{0.00,0.00,1.00}{##1}}}
\expandafter\def\csname PY@tok@gu\endcsname{\let\PY@bf=\textbf\def\PY@tc##1{\textcolor[rgb]{0.50,0.00,0.50}{##1}}}
\expandafter\def\csname PY@tok@sh\endcsname{\def\PY@tc##1{\textcolor[rgb]{0.73,0.13,0.13}{##1}}}
\expandafter\def\csname PY@tok@ch\endcsname{\let\PY@it=\textit\def\PY@tc##1{\textcolor[rgb]{0.25,0.50,0.50}{##1}}}
\expandafter\def\csname PY@tok@cpf\endcsname{\let\PY@it=\textit\def\PY@tc##1{\textcolor[rgb]{0.25,0.50,0.50}{##1}}}
\expandafter\def\csname PY@tok@il\endcsname{\def\PY@tc##1{\textcolor[rgb]{0.40,0.40,0.40}{##1}}}
\expandafter\def\csname PY@tok@k\endcsname{\let\PY@bf=\textbf\def\PY@tc##1{\textcolor[rgb]{0.00,0.50,0.00}{##1}}}
\expandafter\def\csname PY@tok@m\endcsname{\def\PY@tc##1{\textcolor[rgb]{0.40,0.40,0.40}{##1}}}
\expandafter\def\csname PY@tok@se\endcsname{\let\PY@bf=\textbf\def\PY@tc##1{\textcolor[rgb]{0.73,0.40,0.13}{##1}}}
\expandafter\def\csname PY@tok@si\endcsname{\let\PY@bf=\textbf\def\PY@tc##1{\textcolor[rgb]{0.73,0.40,0.53}{##1}}}
\expandafter\def\csname PY@tok@s1\endcsname{\def\PY@tc##1{\textcolor[rgb]{0.73,0.13,0.13}{##1}}}
\expandafter\def\csname PY@tok@kp\endcsname{\def\PY@tc##1{\textcolor[rgb]{0.00,0.50,0.00}{##1}}}
\expandafter\def\csname PY@tok@na\endcsname{\def\PY@tc##1{\textcolor[rgb]{0.49,0.56,0.16}{##1}}}
\expandafter\def\csname PY@tok@nb\endcsname{\def\PY@tc##1{\textcolor[rgb]{0.00,0.50,0.00}{##1}}}
\expandafter\def\csname PY@tok@cm\endcsname{\let\PY@it=\textit\def\PY@tc##1{\textcolor[rgb]{0.25,0.50,0.50}{##1}}}
\expandafter\def\csname PY@tok@s\endcsname{\def\PY@tc##1{\textcolor[rgb]{0.73,0.13,0.13}{##1}}}
\expandafter\def\csname PY@tok@w\endcsname{\def\PY@tc##1{\textcolor[rgb]{0.73,0.73,0.73}{##1}}}
\expandafter\def\csname PY@tok@sr\endcsname{\def\PY@tc##1{\textcolor[rgb]{0.73,0.40,0.53}{##1}}}
\expandafter\def\csname PY@tok@gp\endcsname{\let\PY@bf=\textbf\def\PY@tc##1{\textcolor[rgb]{0.00,0.00,0.50}{##1}}}
\expandafter\def\csname PY@tok@nv\endcsname{\def\PY@tc##1{\textcolor[rgb]{0.10,0.09,0.49}{##1}}}
\expandafter\def\csname PY@tok@mf\endcsname{\def\PY@tc##1{\textcolor[rgb]{0.40,0.40,0.40}{##1}}}
\expandafter\def\csname PY@tok@vi\endcsname{\def\PY@tc##1{\textcolor[rgb]{0.10,0.09,0.49}{##1}}}
\expandafter\def\csname PY@tok@ss\endcsname{\def\PY@tc##1{\textcolor[rgb]{0.10,0.09,0.49}{##1}}}
\expandafter\def\csname PY@tok@kr\endcsname{\let\PY@bf=\textbf\def\PY@tc##1{\textcolor[rgb]{0.00,0.50,0.00}{##1}}}
\expandafter\def\csname PY@tok@c1\endcsname{\let\PY@it=\textit\def\PY@tc##1{\textcolor[rgb]{0.25,0.50,0.50}{##1}}}
\expandafter\def\csname PY@tok@gs\endcsname{\let\PY@bf=\textbf}
\expandafter\def\csname PY@tok@sx\endcsname{\def\PY@tc##1{\textcolor[rgb]{0.00,0.50,0.00}{##1}}}
\expandafter\def\csname PY@tok@gh\endcsname{\let\PY@bf=\textbf\def\PY@tc##1{\textcolor[rgb]{0.00,0.00,0.50}{##1}}}
\expandafter\def\csname PY@tok@kd\endcsname{\let\PY@bf=\textbf\def\PY@tc##1{\textcolor[rgb]{0.00,0.50,0.00}{##1}}}
\expandafter\def\csname PY@tok@mo\endcsname{\def\PY@tc##1{\textcolor[rgb]{0.40,0.40,0.40}{##1}}}
\expandafter\def\csname PY@tok@kt\endcsname{\def\PY@tc##1{\textcolor[rgb]{0.69,0.00,0.25}{##1}}}
\expandafter\def\csname PY@tok@nn\endcsname{\let\PY@bf=\textbf\def\PY@tc##1{\textcolor[rgb]{0.00,0.00,1.00}{##1}}}
\expandafter\def\csname PY@tok@nd\endcsname{\def\PY@tc##1{\textcolor[rgb]{0.67,0.13,1.00}{##1}}}
\expandafter\def\csname PY@tok@vm\endcsname{\def\PY@tc##1{\textcolor[rgb]{0.10,0.09,0.49}{##1}}}
\expandafter\def\csname PY@tok@ne\endcsname{\let\PY@bf=\textbf\def\PY@tc##1{\textcolor[rgb]{0.82,0.25,0.23}{##1}}}
\expandafter\def\csname PY@tok@mb\endcsname{\def\PY@tc##1{\textcolor[rgb]{0.40,0.40,0.40}{##1}}}
\expandafter\def\csname PY@tok@go\endcsname{\def\PY@tc##1{\textcolor[rgb]{0.53,0.53,0.53}{##1}}}
\expandafter\def\csname PY@tok@s2\endcsname{\def\PY@tc##1{\textcolor[rgb]{0.73,0.13,0.13}{##1}}}

\makeatother

\usepackage{natbib}
    \definecolor{incolor}{rgb}{0.0, 0.0, 0.5}
    \definecolor{outcolor}{rgb}{0.545, 0.0, 0.0}

    \hypersetup{
      breaklinks=true,  
      colorlinks=true,
      urlcolor=urlcolor,
      linkcolor=linkcolor,
      citecolor=citecolor,
      }
    
    \title{\vspace{2.0cm}Summary Analysis of the 2017 GitHub Open Source Survey}
    \date{7 June 2017}
\author{\\ R. Stuart Geiger \\ Berkeley Institute for Data Science \\ University of California, Berkeley \\ stuart@stuartgeiger.com}

\begin{document}
    \renewcommand{\footnotesize}{\normalsize} 

    \maketitle
\begin{adjustwidth}{.5in}{.5in}
    \section{Abstract}
This report is a high-level summary analysis of the 2017 GitHub Open Source Survey dataset,\footnote{https://opensourcesurvey.org/2017/} presenting frequency counts, proportions, and frequency or proportion bar plots for every question asked in the survey.
    \section{Overview}
\subsection{The 2017 Open Source Survey}
This report analyzes the open dataset from the 2017 Open Source Survey, which was conducted by staff at GitHub, with help, support, and feedback from many others \citep{Github2017}. The survey was run in 2017, asking over 50 questions on a variety of topics. The survey's designers explain the motivation, design, and distribution of the survey on the project's website: 
\begin{quote}
"In collaboration with researchers from academia, industry, and the community, GitHub designed a survey to gather high quality and novel data on open source software development practices and communities. We collected responses from 5,500 randomly sampled respondents sourced from over 3,800 open source repositories on GitHub.com, and over 500 responses from a non-random sample of communities that work on other platforms. The results are an open data set about the attitudes, experiences, and backgrounds of those who use, build, and maintain open source software." \citep{Github2017-report}
\end{quote}

\subsection{Purpose and goal of this report}

The GitHub survey team presented analyses of some questions when releasing the survey \citep{Github2017-report}, but there were many more questions asked that are relevant to researchers and community members. This report is an exploratory analysis of all questions asked in the survey, providing a basic summary of the responses to each question. This report presents and plots summary statistics -- mostly frequency counts, proportions, then a frequency or proportion bar graph -- of all questions asked in the survey. Most questions are presented individually, with panel questions grouped together as appropriate. There are no correlations, regressions, or descriptive breakouts between subgroups. Likert-style questions (e.g. Strongly agree <-> strongly disagree) have not been recoded to numerical, scalar values. There are no discussions or interpretations of results. This is left for future work.

The purpose of this report is to facilitate future research on this dataset by giving an overview of the kinds of questions asked in the survey, as well as provide a single, stable reference for citing broad claims in the data. This report is the PDF version of a Jupyter Notebook, which can be run to reproduce the results of the tables and graphs in this report. The Jupyter notebook and a copy of the data is public on GitHub \footnote{https://github.com/staeiou/github-survey-analysis} and the Open Science Framework \footnote{https://osf.io/enrq5/}. Others are encouraged to extend it as they see fit, as this report and the notebooks are licensed CC-BY-4.0.\footnote{https://creativecommons.org/licenses/by/4.0/} The "Out[number]" notes before each table and chart are linked to the Jupyter notebook, so you can easily navigate to the notebook cell where the applicable code can be found. If you find this report useful, please cite both this report \citep{geiger-report} and the original survey \citep{Github2017} as detailed in the bibliography at the end of this report.

\subsection{Software used}

This analysis was conducted in Python \citep{python} version 3.6, using Pandas dataframes \citep{pandas} for data parsing and transformation, SciPy \citep{scipy} and NumPy \citep{numpy} for quantitative computations, and Matplotlib \citep{Matplotlib} and Seaborn \citep{seaborn} for visualization. It was conducted in Jupyter Notebooks \citep{jupyter} using the IPython kernel \citep{ipython}, and nbconvert (also discussed in \citep{jupyter}) was used to convert the notebook into LaTeX for publication in this report. 
    \section{Table of Contents}
    \tableofcontents
    
 \end{adjustwidth}

\clearpage

    \section{Analysis}\label{analysis}

\subsection{Contributor identity}\label{contributor-identity}

    \subsubsection{People participate in open source in different ways.
Which of the following activities do you engage
in?}\label{people-participate-in-open-source-in-different-ways.-which-of-the-following-activities-do-you-engage-in}

PARTICIPATION.TYPE.*

\texttt{\color{outcolor}Out[{\color{outcolor}29}]:}
    
    \centering{\begin{tabular}{lrr}
\toprule
{} &    No &   Yes \\
\midrule
PARTICIPATION.TYPE.FOLLOW           &  1287 &  4742 \\
PARTICIPATION.TYPE.USE.APPLICATIONS &   454 &  5575 \\
PARTICIPATION.TYPE.USE.DEPENDENCIES &   946 &  5083 \\
PARTICIPATION.TYPE.CONTRIBUTE       &  1722 &  4307 \\
PARTICIPATION.TYPE.OTHER            &  5742 &   287 \\
\bottomrule
\end{tabular}
}

\texttt{\color{outcolor}Out[{\color{outcolor}30}]:}
    
    \centering{\begin{tabular}{lr}
\toprule
{} &  percent \\
\midrule
PARTICIPATION.TYPE.OTHER            &    4.76\% \\
PARTICIPATION.TYPE.CONTRIBUTE       &   71.44\% \\
PARTICIPATION.TYPE.FOLLOW           &   78.65\% \\
PARTICIPATION.TYPE.USE.DEPENDENCIES &   84.31\% \\
PARTICIPATION.TYPE.USE.APPLICATIONS &   92.47\% \\
\bottomrule
\end{tabular}
}

    \begin{center}
    \adjustimage{max size={0.9\linewidth}{0.9\paperheight}}{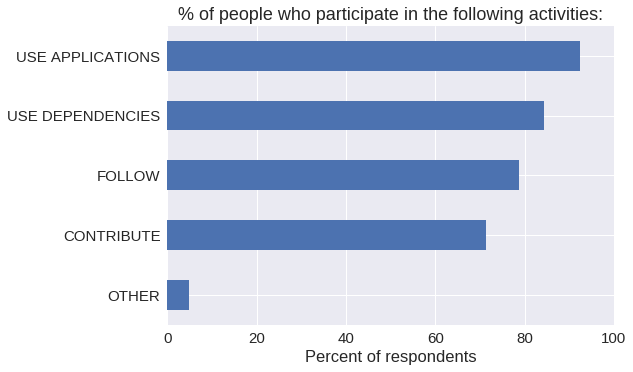}
    \end{center}
    { \hspace*{\fill} \\}
    
    \clearpage     \subsubsection{How often do you engage in each of the following activities?}\label{contributon-type-how-often-do-you-engage-in-each-of-the-following-activities}
    CONTRIBUTION.TYPE.*

\texttt{\color{outcolor}Out[{\color{outcolor}32}]:}
    
    \centering{\begin{tabular}{lrrrr}
\toprule
{} &  Frequently &  Occasionally &  Rarely &  Never \\
\midrule
CONTRIBUTOR.TYPE.COMMUNITY.ADMIN     &         287 &           417 &     867 &   2412 \\
CONTRIBUTOR.TYPE.CONTRIBUTE.DOCS     &         460 &          1214 &    1665 &    661 \\
CONTRIBUTOR.TYPE.FEATURE.REQUESTS    &         573 &          1625 &    1346 &    451 \\
CONTRIBUTOR.TYPE.PROJECT.MAINTENANCE &         996 &           944 &     974 &   1090 \\
CONTRIBUTOR.TYPE.FILE.BUGS           &        1067 &          2073 &     768 &    106 \\
CONTRIBUTOR.TYPE.CONTRIBUTE.CODE     &        1160 &          1383 &    1301 &    189 \\
\bottomrule
\end{tabular}
}

    \begin{center}
    \adjustimage{max size={0.9\linewidth}{0.9\paperheight}}{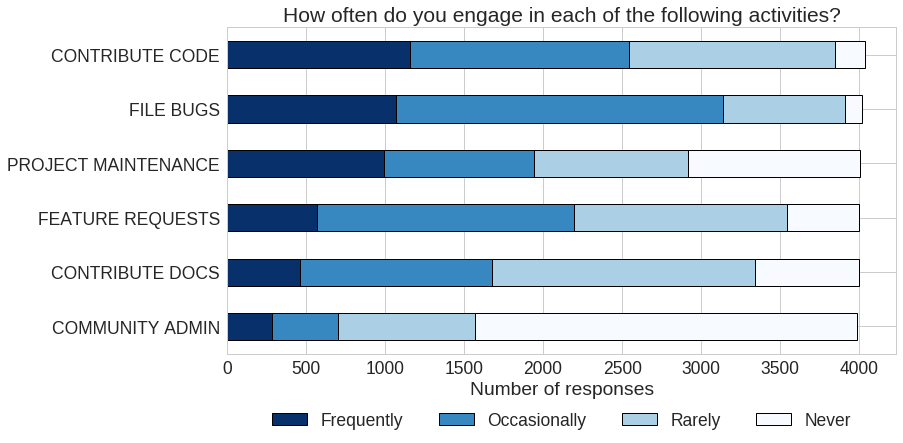}
    \end{center}
    { \hspace*{\fill} \\}
    
    \clearpage     \subsubsection{Employment status}\label{employment-status}

EMPLOYMENT.STATUS

\texttt{\color{outcolor}Out[{\color{outcolor}34}]:}
    
    \centering{\begin{tabular}{lr}
\toprule
{} &  count \\
\midrule
Employed full time                                 &   3615 \\
Full time student                                  &   1048 \\
Employed part time                                 &    349 \\
Temporarily not working                            &    314 \\
Other - please describe                            &    184 \\
Retired or permanently not working (e.g. due to... &     90 \\
\bottomrule
\end{tabular}
}

\texttt{\color{outcolor}Out[{\color{outcolor}35}]:}
    
    \centering{\begin{tabular}{lr}
\toprule
{} &  percent \\
\midrule
Employed full time                                 &   64.55\% \\
Full time student                                  &   18.71\% \\
Employed part time                                 &    6.23\% \\
Temporarily not working                            &    5.61\% \\
Other - please describe                            &    3.29\% \\
Retired or permanently not working (e.g. due to... &    1.61\% \\
\bottomrule
\end{tabular}
}

    \begin{center}
    \adjustimage{max size={0.9\linewidth}{0.9\paperheight}}{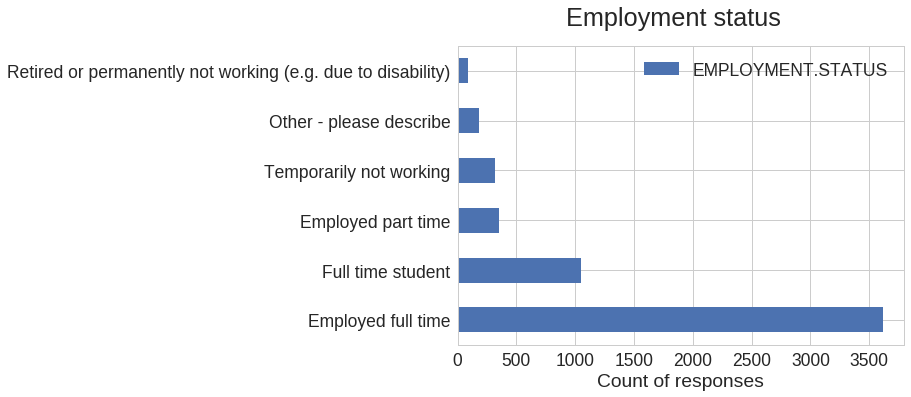}
    \end{center}
    { \hspace*{\fill} \\}
    
    \clearpage     \subsubsection{In your main job, how often do you write or otherwise
directly contribute to producing
software?}\label{in-your-main-job-how-often-do-you-write-or-otherwise-directly-contribute-to-producing-software}

PROFESSIONAL.SOFTWARE

\texttt{\color{outcolor}Out[{\color{outcolor}37}]:}
    
    \centering{\begin{tabular}{lr}
\toprule
{} &  count \\
\midrule
Frequently   &   2747 \\
Occasionally &    542 \\
Rarely       &    339 \\
Never        &    279 \\
\bottomrule
\end{tabular}
}

\texttt{\color{outcolor}Out[{\color{outcolor}38}]:}
    
    \centering{\begin{tabular}{lr}
\toprule
{} &  percent \\
\midrule
Frequently   &   70.31\% \\
Occasionally &   13.87\% \\
Rarely       &    8.68\% \\
Never        &    7.14\% \\
\bottomrule
\end{tabular}
}

    \begin{center}
    \adjustimage{max size={0.9\linewidth}{0.9\paperheight}}{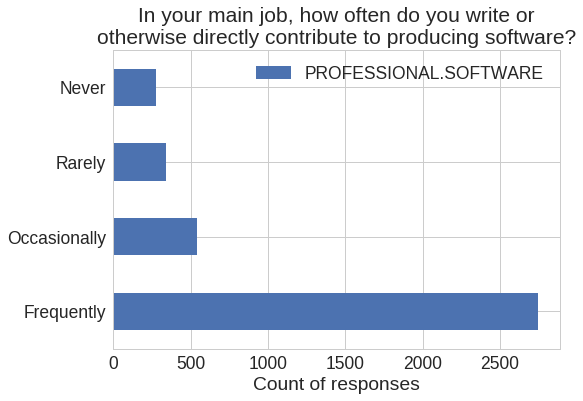}
    \end{center}
    { \hspace*{\fill} \\}
    
    \clearpage     \subsubsection{How interested are you in contributing to open source
projects in the
future?}\label{how-interested-are-you-in-contributing-to-open-source-projects-in-the-future}

FUTURE.CONTRIBUTION.INTEREST

\texttt{\color{outcolor}Out[{\color{outcolor}40}]:}
    
    \centering{\begin{tabular}{lr}
\toprule
{} &  count \\
\midrule
Very interested       &   3929 \\
Somewhat interested   &   1430 \\
Not too interested    &    125 \\
Not at all interested &     24 \\
\bottomrule
\end{tabular}
}

\texttt{\color{outcolor}Out[{\color{outcolor}41}]:}
    
    \centering{\begin{tabular}{lr}
\toprule
{} &  percent \\
\midrule
Very interested       &   71.33\% \\
Somewhat interested   &   25.96\% \\
Not too interested    &    2.27\% \\
Not at all interested &    0.44\% \\
\bottomrule
\end{tabular}
}

    \begin{center}
    \adjustimage{max size={0.9\linewidth}{0.9\paperheight}}{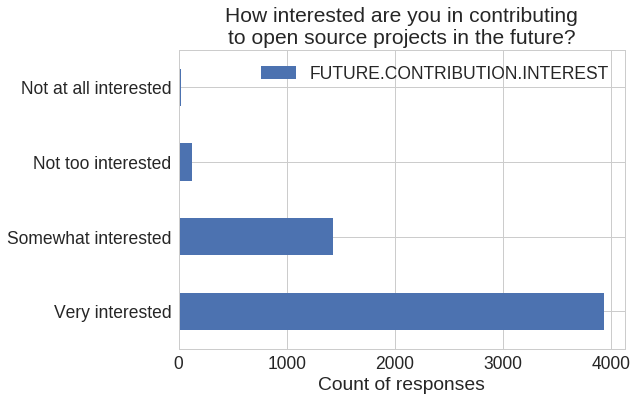}
    \end{center}
    { \hspace*{\fill} \\}
    
    \clearpage     \subsubsection{How likely are you to contribute to open source projects
in the
future?}\label{how-likely-are-you-to-contribute-to-open-source-projects-in-the-future}

\texttt{\color{outcolor}Out[{\color{outcolor}43}]:}
    
    \centering{\begin{tabular}{lr}
\toprule
{} &  count \\
\midrule
Very likely       &   3271 \\
Somewhat likely   &   1719 \\
Somewhat unlikely &    440 \\
Very unlikely     &     81 \\
\bottomrule
\end{tabular}
}

\texttt{\color{outcolor}Out[{\color{outcolor}44}]:}
    
    \centering{\begin{tabular}{lr}
\toprule
{} &  percent \\
\midrule
Very likely       &   59.35\% \\
Somewhat likely   &   31.19\% \\
Somewhat unlikely &    7.98\% \\
Very unlikely     &    1.47\% \\
\bottomrule
\end{tabular}
}

    \begin{center}
    \adjustimage{max size={0.9\linewidth}{0.9\paperheight}}{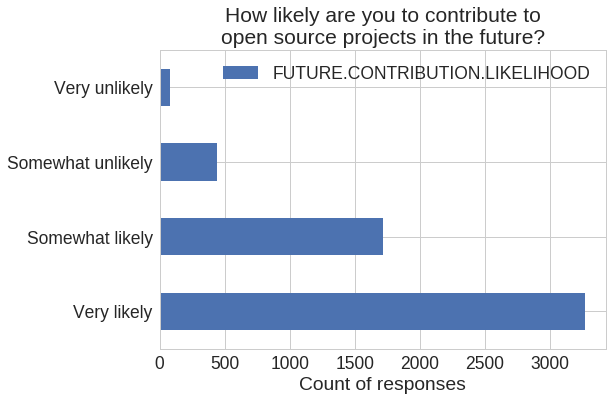}
    \end{center}
    { \hspace*{\fill} \\}
    \clearpage
    \subsection{Priorities and values}\label{priorities-and-values}

     \subsubsection{When thinking about whether to use open source software,
how important are the following
things?}\label{when-thinking-about-whether-to-use-open-source-software-how-important-are-the-following-things}

OSS.USER.PRIORITIES.*

\texttt{\color{outcolor}Out[{\color{outcolor}47}]:}
    
    \centering{\begin{tabular}{lrrrrrr}
\toprule
{} & \pbox{10cm}{Very imp \\ to have} &  \pbox{10cm}{Somewhat \\ imp to have} &  Neither &  \pbox{10cm}{Somewhat \\ imp to not \\ have} &  \pbox{10cm}{Very imp \\ to not have} &  \pbox{10cm}{Don't \\ know} \\
\midrule

CLA                    &                     490 &                        1024 &                      2282 &                             336 &                         157 &                      488 \\
CODE.OF.CONDUCT        &                     848 &                        1461 &                      1993 &                             166 &                         120 &                      209 \\
WIDESPREAD.USE         &                     984 &                        2067 &                      1576 &                             114 &                          47 &                       28 \\
CONTRIBUTING.GUIDE     &                    1212 &                        1866 &                      1516 &                              95 &                          62 &                       62 \\
WELCOMING.COMMUNITY    &                    2062 &                        1822 &                       812 &                              67 &                          33 &                       18 \\
RESPONSIVE.MAINTAINERS &                    2575 &                        1850 &                       302 &                              31 &                          35 &                       20 \\
ACTIVE.DEVELOPMENT     &                    2768 &                        1722 &                       267 &                              30 &                          31 &                       16 \\
LICENSE                &                    3125 &                        1160 &                       435 &                              31 &                          33 &                       47 \\
\bottomrule
\end{tabular}
}

\texttt{\color{outcolor}Out[{\color{outcolor}49}]:}
    
    \centering{\begin{tabular}{lrrrrrr}
\toprule
{} & \pbox{10cm}{Very imp \\ to have} &  \pbox{10cm}{Somewhat \\ imp to have} &  Neither &  \pbox{10cm}{Somewhat \\ imp to not \\ have} &  \pbox{10cm}{Very imp \\ to not have} &  \pbox{10cm}{Don't \\ know} \\
\midrule
CLA                    &                  10.26\% &                      21.44\% &                    47.77\% &                           7.03\% &                       3.29\% &                   10.22\% \\
CODE.OF.CONDUCT        &                  17.68\% &                      30.46\% &                    41.55\% &                           3.46\% &                       2.50\% &                    4.36\% \\
WIDESPREAD.USE         &                  20.43\% &                      42.92\% &                    32.72\% &                           2.37\% &                       0.98\% &                    0.58\% \\
CONTRIBUTING.GUIDE     &                  25.18\% &                      38.77\% &                    31.50\% &                           1.97\% &                       1.29\% &                    1.29\% \\
WELCOMING.COMMUNITY    &                  42.83\% &                      37.85\% &                    16.87\% &                           1.39\% &                       0.69\% &                    0.37\% \\
RESPONSIVE.MAINTAINERS &                  53.50\% &                      38.44\% &                     6.27\% &                           0.64\% &                       0.73\% &                    0.42\% \\
ACTIVE.DEVELOPMENT     &                  57.26\% &                      35.62\% &                     5.52\% &                           0.62\% &                       0.64\% &                    0.33\% \\
LICENSE                &                  64.69\% &                      24.01\% &                     9.00\% &                           0.64\% &                       0.68\% &                    0.97\% \\
\bottomrule
\end{tabular}
}

    \begin{center}
    \adjustimage{max size={0.9\linewidth}{0.9\paperheight}}{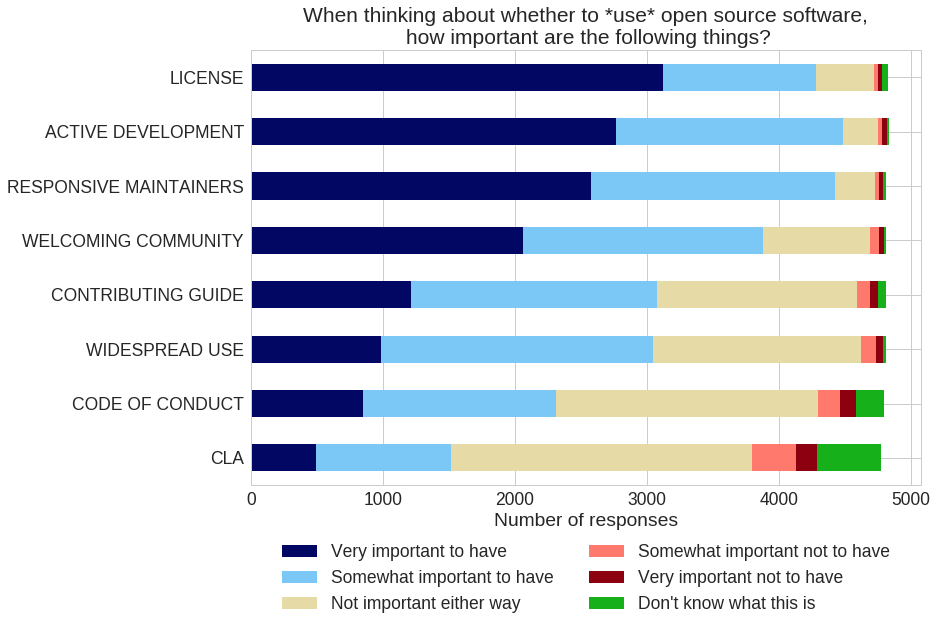}
    \end{center}
    { \hspace*{\fill} \\}
    
    \clearpage     \subsubsection{When thinking about whether to contribute to an open
source project, how important are the following
things?}\label{when-thinking-about-whether-to-contribute-to-an-open-source-project-how-important-are-the-following-things}

OSS.CONTRIBUTOR.PRIORITIES.*

\texttt{\color{outcolor}Out[{\color{outcolor}52}]:}
    
    \centering{\begin{tabular}{lrrrrrr}
\toprule
{} & \pbox{10cm}{Very imp \\ to have} &  \pbox{10cm}{Somewhat \\ imp to have} &  Neither &  \pbox{10cm}{Somewhat \\ imp to not \\ have} &  \pbox{10cm}{Very imp \\ to not have} &  \pbox{10cm}{Don't \\ know} \\
\midrule
WIDESPREAD.USE         &                     387 &                        1016 &                      1666 &                              70 &                          30 &                       12 \\
CLA                    &                     419 &                         712 &                      1266 &                             327 &                         166 &                      280 \\
CODE.OF.CONDUCT        &                     655 &                        1145 &                      1085 &                             119 &                          84 &                       96 \\
CONTRIBUTING.GUIDE     &                    1198 &                        1396 &                       500 &                              41 &                          18 &                       24 \\
ACTIVE.DEVELOPMENT     &                    1368 &                        1333 &                       448 &                              21 &                          18 &                        5 \\
WELCOMING.COMMUNITY    &                    1533 &                        1199 &                       411 &                              21 &                          15 &                        7 \\
RESPONSIVE.MAINTAINERS &                    1994 &                        1022 &                       138 &                               7 &                          16 &                        7 \\
LICENSE                &                    2199 &                         610 &                       337 &                              16 &                          15 &                       18 \\
\bottomrule
\end{tabular}
}

\texttt{\color{outcolor}Out[{\color{outcolor}54}]:}
    
    \centering{\begin{tabular}{lrrrrrr}
\toprule
{} & \pbox{10cm}{Very imp \\ to have} &  \pbox{10cm}{Somewhat \\ imp to have} &  Neither &  \pbox{10cm}{Somewhat \\ imp to not \\ have} &  \pbox{10cm}{Very imp \\ to not have} &  \pbox{10cm}{Don't \\ know} \\
\midrule
WIDESPREAD.USE         &                  12.17\% &                      31.94\% &                    52.37\% &                           2.20\% &                       0.94\% &                    0.38\% \\
CLA                    &                  13.22\% &                      22.46\% &                    39.94\% &                          10.32\% &                       5.24\% &                    8.83\% \\
CODE.OF.CONDUCT        &                  20.57\% &                      35.96\% &                    34.08\% &                           3.74\% &                       2.64\% &                    3.02\% \\
CONTRIBUTING.GUIDE     &                  37.71\% &                      43.94\% &                    15.74\% &                           1.29\% &                       0.57\% &                    0.76\% \\
ACTIVE.DEVELOPMENT     &                  42.84\% &                      41.75\% &                    14.03\% &                           0.66\% &                       0.56\% &                    0.16\% \\
WELCOMING.COMMUNITY    &                  48.12\% &                      37.63\% &                    12.90\% &                           0.66\% &                       0.47\% &                    0.22\% \\
RESPONSIVE.MAINTAINERS &                  62.63\% &                      32.10\% &                     4.33\% &                           0.22\% &                       0.50\% &                    0.22\% \\
LICENSE                &                  68.83\% &                      19.09\% &                    10.55\% &                           0.50\% &                       0.47\% &                    0.56\% \\
\bottomrule
\end{tabular}
}

    \begin{center}
    \adjustimage{max size={0.9\linewidth}{0.9\paperheight}}{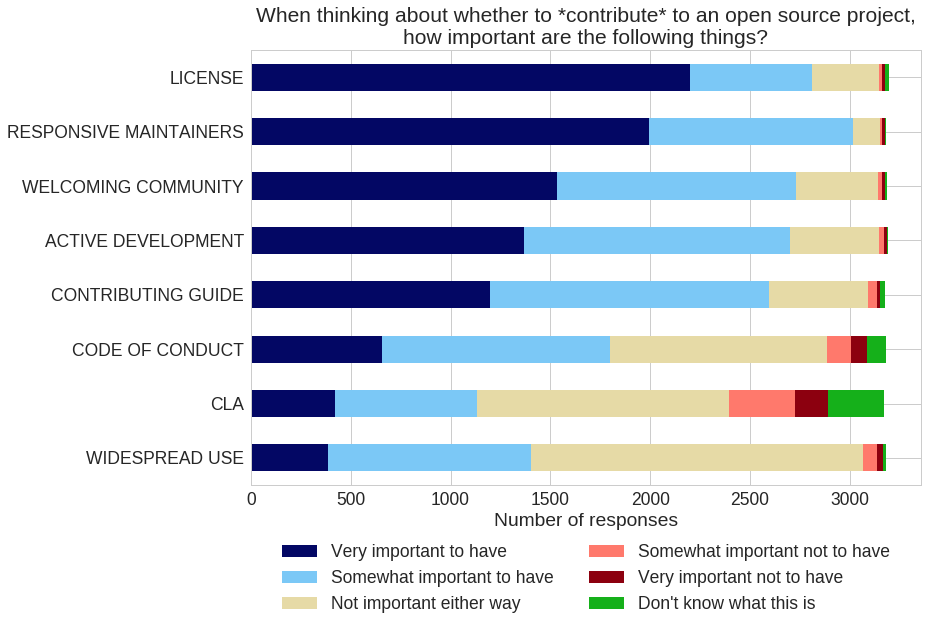}
    \end{center}
    { \hspace*{\fill} \\}
    
    \clearpage     \subsubsection{How often do you try to find open source options over
other kinds of
software?}\label{how-often-do-you-try-to-find-open-source-options-over-other-kinds-of-software}

SEEK.OPEN.SOURCE

\texttt{\color{outcolor}Out[{\color{outcolor}56}]:}
    
    \centering{\begin{tabular}{lr}
\toprule
{} &  count \\
\midrule
Always    &   3407 \\
Sometimes &   1111 \\
Rarely    &    100 \\
Never     &     25 \\
\bottomrule
\end{tabular}
}

\texttt{\color{outcolor}Out[{\color{outcolor}57}]:}
    
    \centering{\begin{tabular}{lr}
\toprule
{} &  percent \\
\midrule
Always    &   73.38\% \\
Sometimes &   23.93\% \\
Rarely    &    2.15\% \\
Never     &    0.54\% \\
\bottomrule
\end{tabular}
}

    \begin{center}
    \adjustimage{max size={0.9\linewidth}{0.9\paperheight}}{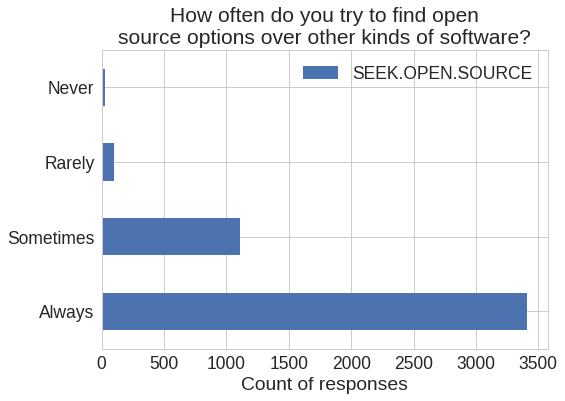}
    \end{center}
    { \hspace*{\fill} \\}
    
    \clearpage     \subsubsection{Open source software
usability}\label{open-source-software-usability}

OSS.UX: Do you believe that open source software is generally easier to
use than closed source (proprietary) software, harder to use, or about
the same?

\texttt{\color{outcolor}Out[{\color{outcolor}59}]:}
    
    \centering{\begin{tabular}{lr}
\toprule
{} &  count \\
\midrule
About the same          &   2027 \\
Generally easier to use &   1597 \\
Generally harder to use &    897 \\
\bottomrule
\end{tabular}
}

\texttt{\color{outcolor}Out[{\color{outcolor}60}]:}
    
    \centering{\begin{tabular}{lr}
\toprule
{} &  percent \\
\midrule
About the same          &   44.84\% \\
Generally easier to use &   35.32\% \\
Generally harder to use &   19.84\% \\
\bottomrule
\end{tabular}
}

    \begin{center}
    \adjustimage{max size={0.9\linewidth}{0.9\paperheight}}{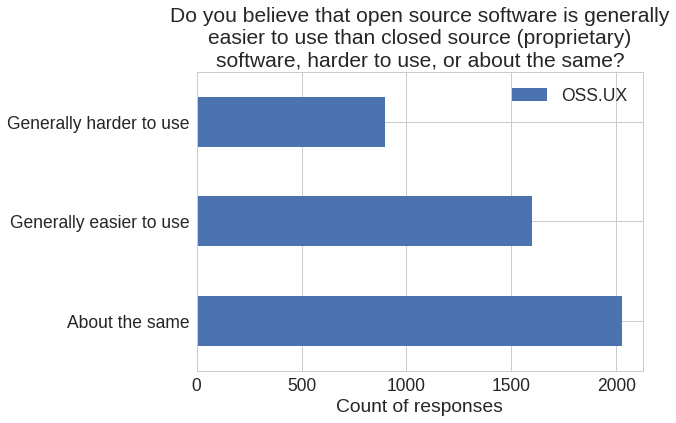}
    \end{center}
    { \hspace*{\fill} \\}
    
    \clearpage     \subsubsection{Open source software
security}\label{open-source-software-security}

OSS.SECURITY: Do you believe that open source software is generally more
secure than closed source (proprietary) software, less secure, or about
the same?

\texttt{\color{outcolor}Out[{\color{outcolor}62}]:}
    
    \centering{\begin{tabular}{lr}
\toprule
{} &  count \\
\midrule
Generally more secure &   2688 \\
About the same        &   1537 \\
Generally less secure &    295 \\
\bottomrule
\end{tabular}
}

\texttt{\color{outcolor}Out[{\color{outcolor}63}]:}
    
    \centering{\begin{tabular}{lr}
\toprule
{} &  percent \\
\midrule
Generally more secure &   59.47\% \\
About the same        &   34.00\% \\
Generally less secure &    6.53\% \\
\bottomrule
\end{tabular}
}

    \begin{center}
    \adjustimage{max size={0.9\linewidth}{0.9\paperheight}}{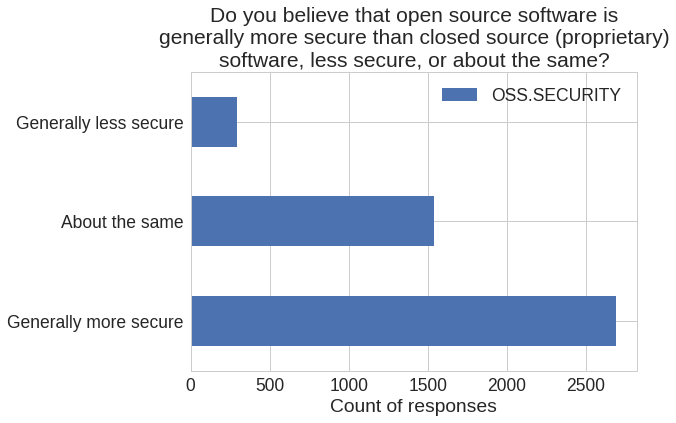}
    \end{center}
    { \hspace*{\fill} \\}
    
    \clearpage     \subsubsection{Open source software
stability}\label{open-source-software-stability}

OSS.STABILITY: Do you believe that open source software is generally
more stable than closed source (proprietary) software, less stable, or
about the same?

\texttt{\color{outcolor}Out[{\color{outcolor}65}]:}
    
    \centering{\begin{tabular}{lr}
\toprule
{} &  count \\
\midrule
About the same        &   2240 \\
Generally more stable &   1399 \\
Generally less stable &    877 \\
\bottomrule
\end{tabular}
}

\texttt{\color{outcolor}Out[{\color{outcolor}66}]:}
    
    \centering{\begin{tabular}{lr}
\toprule
{} &  percent \\
\midrule
About the same        &   49.60\% \\
Generally more stable &   30.98\% \\
Generally less stable &   19.42\% \\
\bottomrule
\end{tabular}
}

    \begin{center}
    \adjustimage{max size={0.9\linewidth}{0.9\paperheight}}{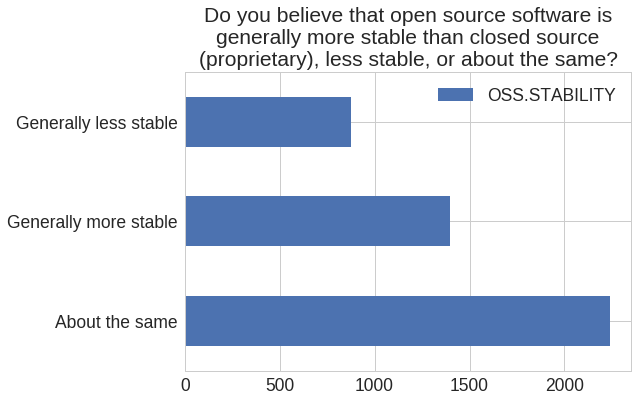}
    \end{center}
    { \hspace*{\fill} \\}

    \clearpage     \subsubsection{Identification with open
source}\label{identification-with-open-source}

How much do you agree or disagree with the following statements:

\begin{itemize}
\tightlist
\item
  EXTERNAL.EFFICACY: The open source community values contributions from
  people like me.
\item
  INTERNAL.EFFICACY: I have the skills and understanding necessary to
  make meaningful contributions to open source projects.
\item
  OSS.IDENTIFICATION: I consider myself to be a member of the open
  source (and/or the Free/Libre software) community.
\end{itemize}

\texttt{\color{outcolor}Out[{\color{outcolor}69}]:}
    
    \centering{\begin{tabular}{lrrrrr}
\toprule
{} &  \pbox{10cm}{Strongly agree} &  \pbox{10cm}{Somewhat agree} &  Neither &  \pbox{10cm}{Somewhat disagree} &  \pbox{10cm}{Strongly disagree} \\
\midrule
EXTERNAL.EFFICACY  &            1518 &            1610 &                        1116 &                150 &                 58 \\
OSS.IDENTIFICATION &            1579 &            1513 &                         863 &                351 &                150 \\
INTERNAL.EFFICACY  &            2052 &            1685 &                         418 &                240 &                 62 \\
\bottomrule
\end{tabular}
}

\texttt{\color{outcolor}Out[{\color{outcolor}70}]:}
    
    \centering{\begin{tabular}{lrrrrr}
\toprule
{} &  \pbox{10cm}{Strongly agree} &  \pbox{10cm}{Somewhat agree} &  Neither &  \pbox{10cm}{Somewhat disagree} &  \pbox{10cm}{Strongly disagree} \\
\midrule
INTERNAL.EFFICACY  &                       9.38\% &          37.81\% &              5.38\% &          46.04\% &              1.39\% \\
EXTERNAL.EFFICACY  &                      25.07\% &          36.16\% &              3.37\% &          34.10\% &              1.30\% \\
OSS.IDENTIFICATION &                      19.37\% &          33.95\% &              7.88\% &          35.44\% &              3.37\% \\
\bottomrule
\end{tabular}
}

    \begin{center}
    \adjustimage{max size={1\linewidth}{0.9\paperheight}}{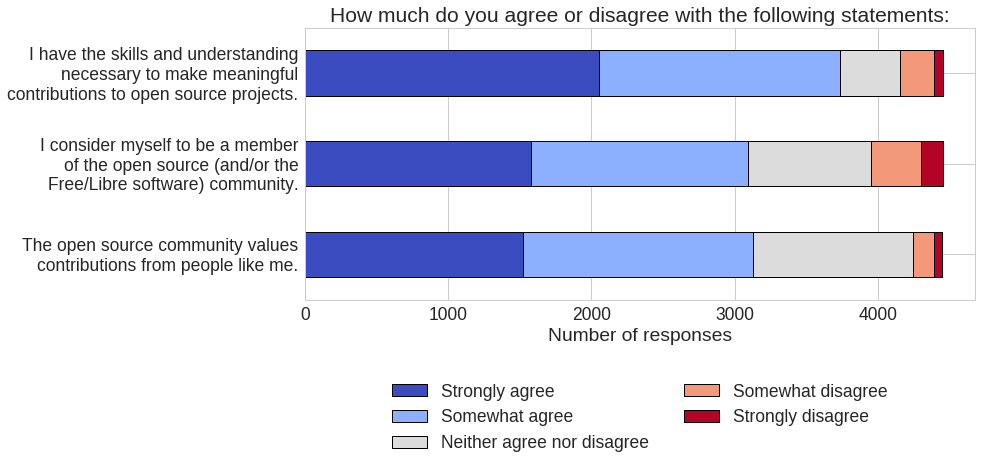}
    \end{center}
    { \hspace*{\fill} \\}
    \clearpage
    \subsection{Transparency vs privacy}\label{transparency-vs-privacy}

    \subsubsection{Attribution}\label{attribution}

TRANSPARENCY.PRIVACY.BELIEFS: Which of the following statements is
closest to your beliefs about attribution in software development?

\begin{itemize}
\tightlist
\item
  Records of authorship should be required so that end users know who
  created the source code they are working with.
\item
  People should be able to contribute code without attribution, if they
  wish to remain anonymous.
\end{itemize}

\texttt{\color{outcolor}Out[{\color{outcolor}72}]:}
    
    \centering{\begin{tabular}{lr}
\toprule
{} &  count \\
\midrule
People should be able to contribute code withou... &   2454 \\
Records of authorship should be required so tha... &   1594 \\
\bottomrule
\end{tabular}
}

\texttt{\color{outcolor}Out[{\color{outcolor}73}]:}
    
    \centering{\begin{tabular}{lr}
\toprule
{} &  percent \\
\midrule
People should be able to contribute code withou... &   60.62\% \\
Records of authorship should be required so tha... &   39.38\% \\
\bottomrule
\end{tabular}
}

    \begin{center}
    \adjustimage{max size={0.9\linewidth}{0.9\paperheight}}{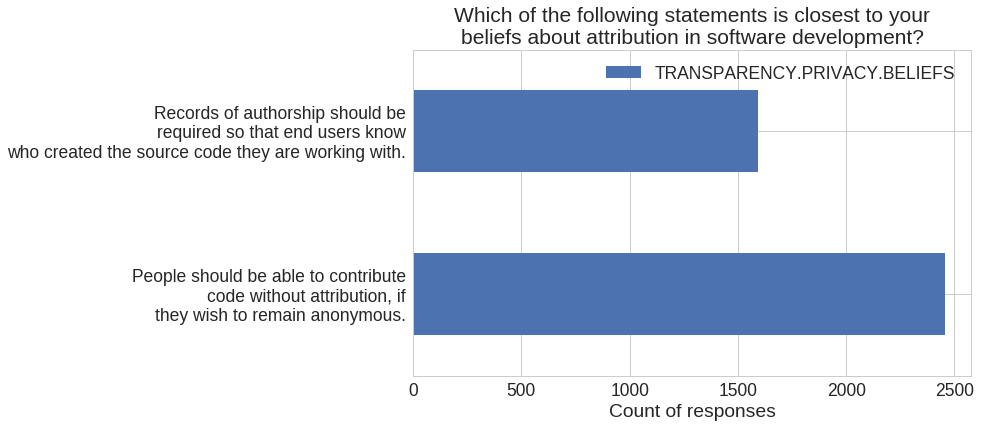}
    \end{center}
    { \hspace*{\fill} \\}
    
    \clearpage     \subsubsection{In general, how much information about you is publicly
available
online?}\label{in-general-how-much-information-about-you-is-publicly-available-online}

INFO.AVAILABILITY

\texttt{\color{outcolor}Out[{\color{outcolor}75}]:}
    
    \centering{\begin{tabular}{lr}
\toprule
{} &  count \\
\midrule
Some information about me      &   1776 \\
A little information about me  &   1133 \\
A lot of information about me  &   1011 \\
No information at all about me &    140 \\
\bottomrule
\end{tabular}
}

\texttt{\color{outcolor}Out[{\color{outcolor}76}]:}
    
    \centering{\begin{tabular}{lr}
\toprule
{} &  percent \\
\midrule
Some information about me      &   43.74\% \\
A little information about me  &   27.91\% \\
A lot of information about me  &   24.90\% \\
No information at all about me &    3.45\% \\
\bottomrule
\end{tabular}
}

    \begin{center}
    \adjustimage{max size={0.9\linewidth}{0.9\paperheight}}{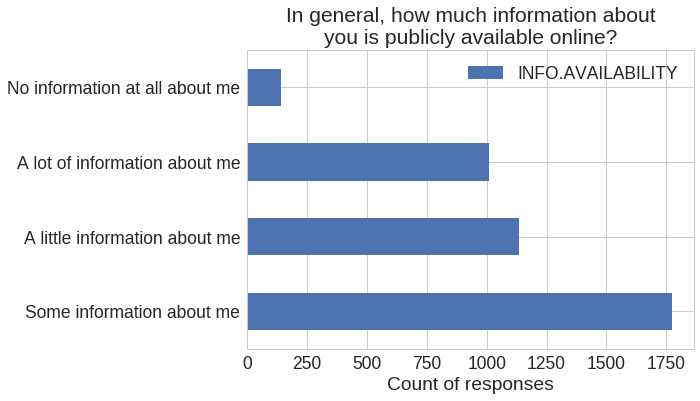}
    \end{center}
    { \hspace*{\fill} \\}
    
    \clearpage     \subsubsection{Do you feel that you need to make information available
about yourself online for professional
reasons?}\label{do-you-feel-that-you-need-to-make-information-available-about-yourself-online-for-professional-reasons}

INFO.JOB

\texttt{\color{outcolor}Out[{\color{outcolor}78}]:}
    
    \centering{\begin{tabular}{lr}
\toprule
{} &  count \\
\midrule
Yes &   2327 \\
No  &   1638 \\
\bottomrule
\end{tabular}
}

\texttt{\color{outcolor}Out[{\color{outcolor}79}]:}
    
    \centering{\begin{tabular}{lr}
\toprule
{} &  percent \\
\midrule
Yes &   58.69\% \\
No  &   41.31\% \\
\bottomrule
\end{tabular}
}

    \begin{center}
    \adjustimage{max size={0.6\linewidth}{0.9\paperheight}}{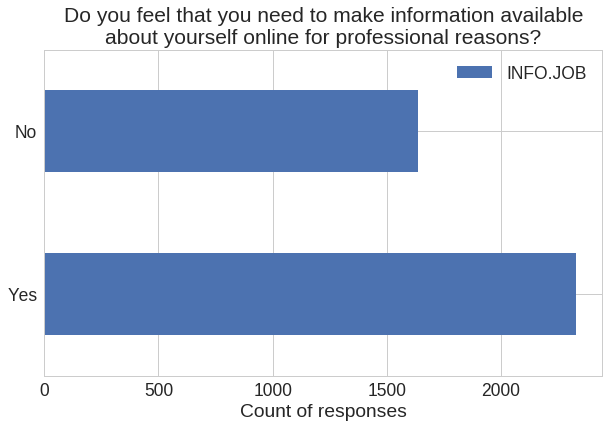}
    \end{center}
    { \hspace*{\fill} \\}
    
    \clearpage     \subsubsection{General privacy
practices}\label{general-privacy-practices}

TRANSPARENCY.PRIVACY.PRACTICES.GENERAL

"Which of the following best describes your practices around publishing
content online, such as posts on social media (e.g. Facebook, Instagram,
Twitter, etc.), blogs, and other platforms (not including contributions
to open source projects)?" (single choice)

\texttt{\color{outcolor}Out[{\color{outcolor}81}]:}
    
    \centering{\begin{tabular}{lr}
\toprule
{} &  count \\
\midrule
I include my real name.                            &   1718 \\
I usually use a consistent pseudonym that is easily linked to my real name online. &   1141 \\
I don't publish this kind of content online.       &    517 \\
I usually use a consistent pseudonym that is not linked anywhere with my real name online &    363 \\
I take precautions to use different pseudonymns on different platforms. &    270 \\
\bottomrule
\end{tabular}
}

\texttt{\color{outcolor}Out[{\color{outcolor}82}]:}
    
    \centering{\begin{tabular}{lr}
\toprule
{} &  percent \\
\midrule
I include my real name.                            &   42.85\% \\
I usually use a consistent pseudonym that is easily linked to my real name online. &   28.46\% \\
I don't publish this kind of content online.       &   12.90\% \\
I usually use a consistent pseudonym that is not linked anywhere with my real name online &    9.05\% \\
I take precautions to use different pseudonymns on different platforms. &    6.73\% \\
\bottomrule
\end{tabular}
}

    \begin{center}
    \adjustimage{max size={1\linewidth}{0.9\paperheight}}{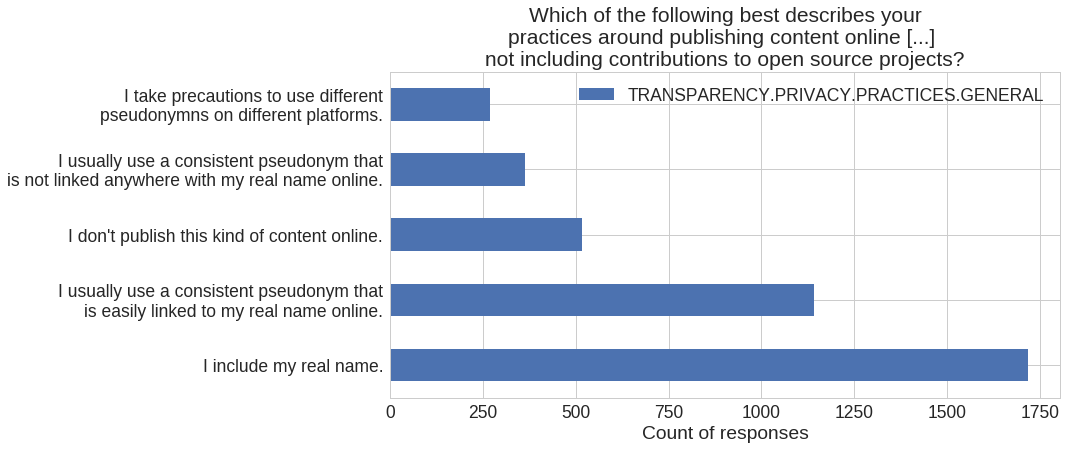}
    \end{center}
    { \hspace*{\fill} \\}
    
    \clearpage     \subsubsection{OSS privacy practices}\label{oss-privacy-practices}

"Which of the following best describes your practices when making open
source contributions?"

\texttt{\color{outcolor}Out[{\color{outcolor}85}]:}
    
    \centering{\begin{tabular}{lr}
\toprule
{} &  count \\
\midrule
I include my real name.                            &   1845 \\
I usually use a consistent pseudonym that is easily linked to my real name online. &    766 \\
I usually use a consistent pseudonym that is not linked anywhere with my real name online &    273 \\
I take precautions to use different pseudonymns on different platforms. &     42 \\
\bottomrule
\end{tabular}
}

\texttt{\color{outcolor}Out[{\color{outcolor}86}]:}
    
    \centering{\begin{tabular}{lr}
\toprule
{} &  percent \\
\midrule
I include my real name.                            &   63.06\% \\
I usually use a consistent pseudonym that is easily linked to my real name online. &   26.18\% \\
I usually use a consistent pseudonym that is not linked anywhere with my real name online &    9.33\% \\
I take precautions to use different pseudonymns on different platforms. &    1.44\% \\
\bottomrule
\end{tabular}
}

    \begin{center}
    \adjustimage{max size={1\linewidth}{0.9\paperheight}}{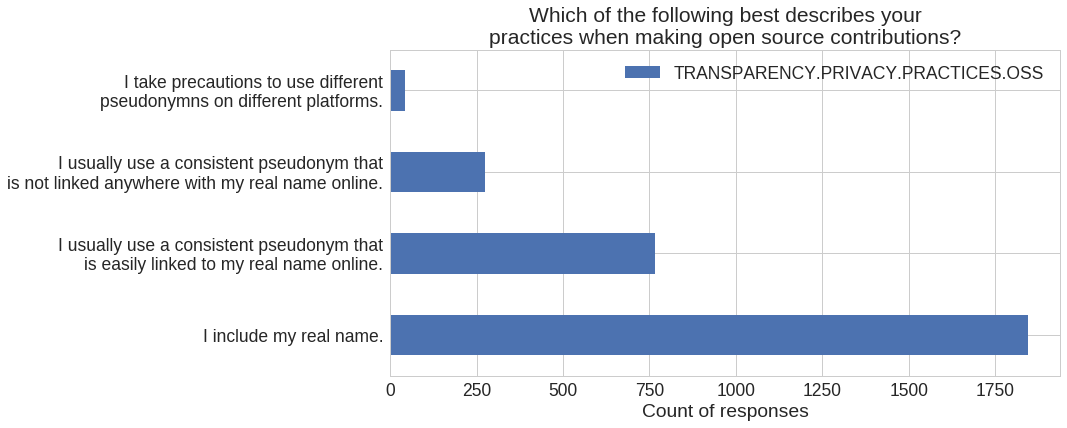}
    \end{center}
    { \hspace*{\fill} \\}
    \clearpage 
    \subsection{Mentorship / Help}\label{mentorship-help}

    \subsubsection{Have you ever received any kind of help from other people
related to using or contributing to an open source
project?}\label{have-you-ever-received-any-kind-of-help-from-other-people-related-to-using-or-contributing-to-an-open-source-project}

RECEIVED.HELP

\texttt{\color{outcolor}Out[{\color{outcolor}89}]:}
    
    \centering{\begin{tabular}{lr}
\toprule
{} &  count \\
\midrule
Yes &   2845 \\
No  &   1064 \\
\bottomrule
\end{tabular}
}

\texttt{\color{outcolor}Out[{\color{outcolor}90}]:}
    
    \centering{\begin{tabular}{lr}
\toprule
{} &  percent \\
\midrule
Yes &   72.78\% \\
No  &   27.22\% \\
\bottomrule
\end{tabular}
}

    \begin{center}
    \adjustimage{max size={0.7\linewidth}{0.9\paperheight}}{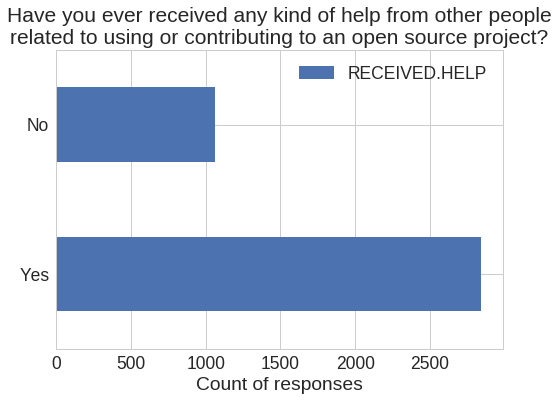}
    \end{center}
    { \hspace*{\fill} \\}
    
    \clearpage     \subsubsection{Thinking of the most recent case where someone helped
you, how did you find someone to help
you?}\label{thinking-of-the-most-recent-case-where-someone-helped-you-how-did-you-find-someone-to-help-you}

\texttt{\color{outcolor}Out[{\color{outcolor}92}]:}
    
    \centering{\begin{tabular}{lr}
\toprule
{} &  count \\
\midrule
I asked for help in a public forum &   2057 \\
I asked a specific person for help.                &    403 \\
Someone offered me unsolicited help.               &    272 \\
Other - Please describe                            &     64 \\
\bottomrule
\end{tabular}
}

\texttt{\color{outcolor}Out[{\color{outcolor}93}]:}
    
    \centering{\begin{tabular}{lr}
\toprule
{} &  percent \\
\midrule
I asked for help in a public forum  &   73.57\% \\
I asked a specific person for help.                &   14.41\% \\
Someone offered me unsolicited help.               &    9.73\% \\
Other - Please describe                            &    2.29\% \\
\bottomrule
\end{tabular}
}

    \begin{center}
    \adjustimage{max size={0.9\linewidth}{0.9\paperheight}}{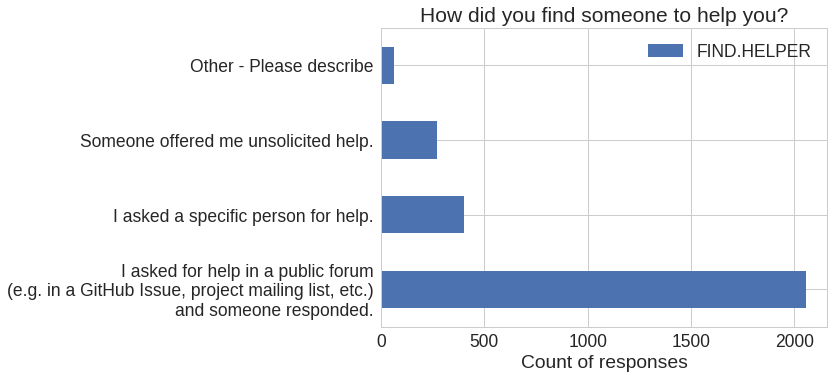}
    \end{center}
    { \hspace*{\fill} \\}
    
    \clearpage     \subsubsection{Which best describes your prior relationship with the
person who helped
you?}\label{which-best-describes-your-prior-relationship-with-the-person-who-helped-you}

HELPER.PRIOR.RELATIONSHIP

\texttt{\color{outcolor}Out[{\color{outcolor}95}]:}
    
    \centering{\begin{tabular}{lr}
\toprule
{} &  count \\
\midrule
Total strangers, I didn't know of them previously. &   1565 \\
I knew of them through their contributions to projects, but didn't know them personally. &    809 \\
We knew each other a little.                       &    211 \\
We knew each other well.                           &    208 \\
\bottomrule
\end{tabular}
}

\texttt{\color{outcolor}Out[{\color{outcolor}96}]:}
    
    \centering{\begin{tabular}{lr}
\toprule
{} &  percent \\
\midrule
Total strangers, I didn't know of them previously. &   56.03\% \\
I knew of them through their contributions to projects, but didn't know them personally. &   28.97\% \\
We knew each other a little.                       &    7.55\% \\
We knew each other well.                           &    7.45\% \\
\bottomrule
\end{tabular}
}

    \begin{center}
    \adjustimage{max size={1\linewidth}{0.9\paperheight}}{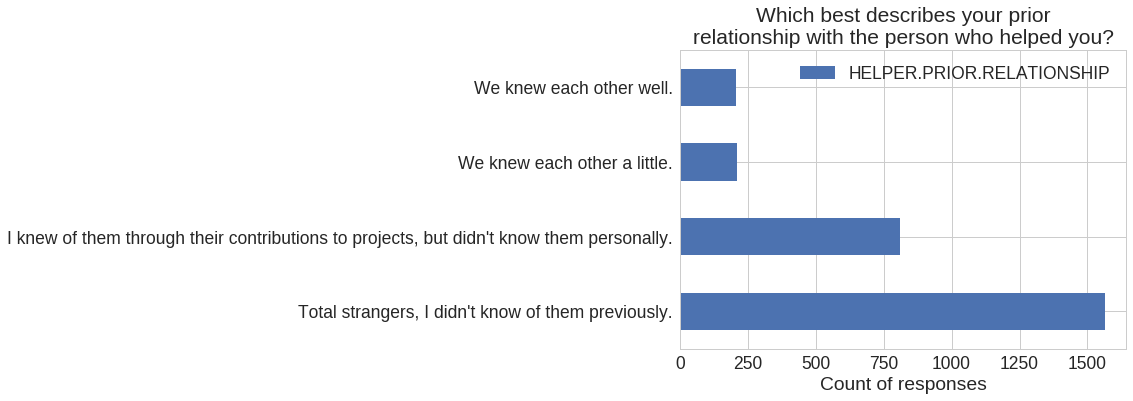}
    \end{center}
    { \hspace*{\fill} \\}
    
    \clearpage     \subsubsection{What kind of problem did they help you
with?}\label{what-kind-of-problem-did-they-help-you-with}

RECEIVED.HELP.TYPE

\texttt{\color{outcolor}Out[{\color{outcolor}98}]:}
    
    \centering{\begin{tabular}{lr}
\toprule
{} &  count \\
\midrule
Writing code or otherwise implementing ideas.      &   1633 \\
Installing or using an application.                &    820 \\
Understanding community norms &    181 \\
Other (please describe)                            &    142 \\
Introductions to other people                      &     13 \\
\bottomrule
\end{tabular}
}

\texttt{\color{outcolor}Out[{\color{outcolor}99}]:}
    
    \centering{\begin{tabular}{lr}
\toprule
{} &  percent \\
\midrule
Writing code or otherwise implementing ideas.      &   58.55\% \\
Installing or using an application.                &   29.40\% \\
Understanding community norms &    6.49\% \\
Other (please describe)                            &    5.09\% \\
Introductions to other people                      &    0.47\% \\
\bottomrule
\end{tabular}
}

    \begin{center}
    \adjustimage{max size={0.9\linewidth}{0.9\paperheight}}{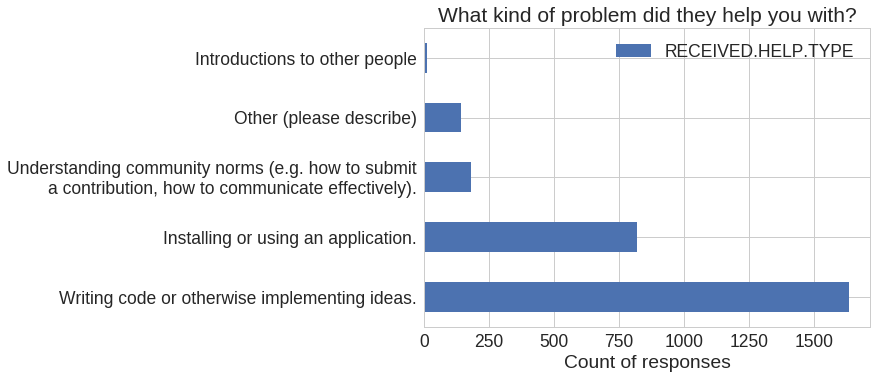}
    \end{center}
    { \hspace*{\fill} \\}
    
    \clearpage     \subsubsection{Have you ever provided help for another person on an open
source
project?}\label{have-you-ever-provided-help-for-another-person-on-an-open-source-project}

PROVIDED.HELP

\texttt{\color{outcolor}Out[{\color{outcolor}101}]:}
    
    \centering{\begin{tabular}{lr}
\toprule
{} &  count \\
\midrule
Yes &   2891 \\
No  &   1013 \\
\bottomrule
\end{tabular}
}

\texttt{\color{outcolor}Out[{\color{outcolor}102}]:}
    
    \centering{\begin{tabular}{lr}
\toprule
{} &  percent \\
\midrule
Yes &   74.05\% \\
No  &   25.95\% \\
\bottomrule
\end{tabular}
}

    \begin{center}
    \adjustimage{max size={0.7\linewidth}{0.9\paperheight}}{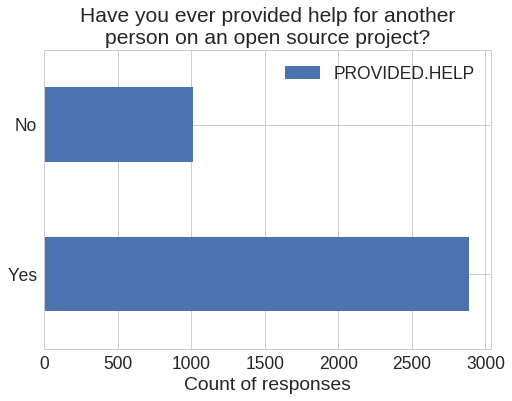}
    \end{center}
    { \hspace*{\fill} \\}

    \clearpage     \subsubsection{Thinking of the most recent case where you helped
someone, how did you come to help this
person?}\label{thinking-of-the-most-recent-case-where-you-helped-someone-how-did-you-come-to-help-this-person}

FIND.HELPEE

\texttt{\color{outcolor}Out[{\color{outcolor}104}]:}
    
    \centering{\begin{tabular}{lr}
\toprule
{} &  count \\
\midrule
They asked for help in a public forum &   1839 \\
They asked me directly for help.                   &    566 \\
I reached out to them to offer unsolicited help.   &    405 \\
Other (please describe)                            &     28 \\
\bottomrule
\end{tabular}
}

\texttt{\color{outcolor}Out[{\color{outcolor}105}]:}
    
    \centering{\begin{tabular}{lr}
\toprule
{} &  percent \\
\midrule
They asked for help in a public forum  &   64.80\% \\
They asked me directly for help.                   &   19.94\% \\
I reached out to them to offer unsolicited help.   &   14.27\% \\
Other (please describe)                            &    0.99\% \\
\bottomrule
\end{tabular}
}

    \begin{center}
    \adjustimage{max size={0.9\linewidth}{0.9\paperheight}}{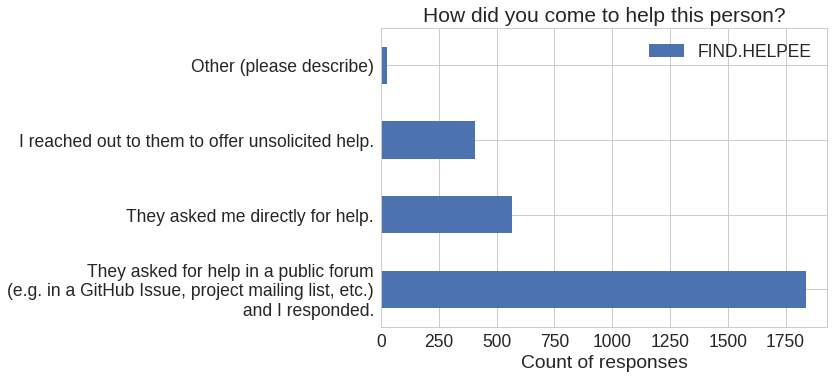}
    \end{center}
    { \hspace*{\fill} \\}
    
    \clearpage     \subsubsection{Which best describes your prior relationship with the
person you
helped?}\label{which-best-describes-your-prior-relationship-with-the-person-you-helped}

HELPEE.PRIOR.RELATIONSHIP

\texttt{\color{outcolor}Out[{\color{outcolor}107}]:}
    
    \centering{\begin{tabular}{lr}
\toprule
{} &  count \\
\midrule
Total strangers, I didn't know of them previously. &   1984 \\
We knew each other well.                           &    292 \\
I knew of them through their contributions to projects, but didn't know them personally. &    288 \\
We knew each other  a little.                      &    275 \\
\bottomrule
\end{tabular}
}

\texttt{\color{outcolor}Out[{\color{outcolor}108}]:}
    
    \centering{\begin{tabular}{lr}
\toprule
{} &  percent \\
\midrule
Total strangers, I didn't know of them previously. &   69.88\% \\
We knew each other well.                           &   10.29\% \\
I knew of them through their contributions to projects, but didn't know them personally. &   10.14\% \\
We knew each other  a little.                      &    9.69\% \\
\bottomrule
\end{tabular}
}

    \begin{center}
    \adjustimage{max size={1\linewidth}{0.9\paperheight}}{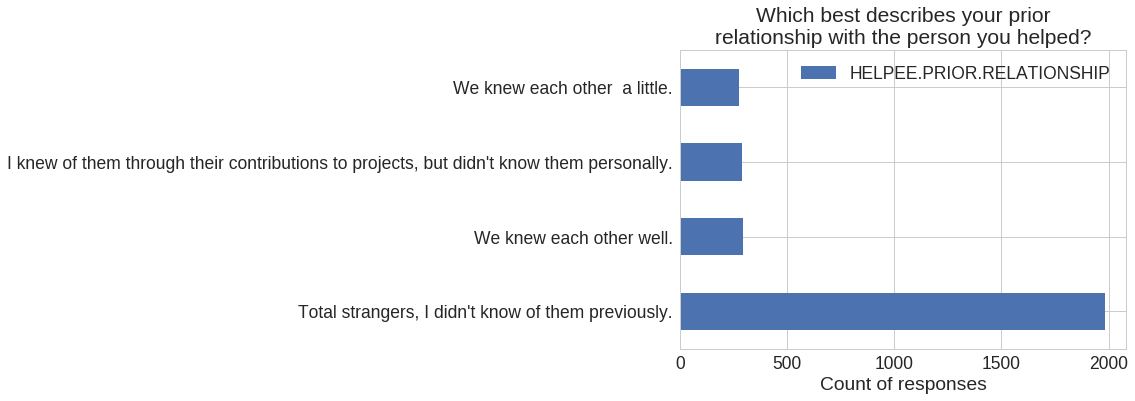}
    \end{center}
    { \hspace*{\fill} \\}
    
    \clearpage     \subsubsection{What kind of problem did you help them
with?}\label{what-kind-of-problem-did-you-help-them-with}

PROVIDED.HELP.TYPE

\texttt{\color{outcolor}Out[{\color{outcolor}110}]:}
    
    \centering{\begin{tabular}{lr}
\toprule
{} &  count \\
\midrule
Writing code or otherwise implementing ideas.      &   1602 \\
Installing or using an application.                &   1028 \\
Other (please describe)                            &    101 \\
Understanding community norms  &     99 \\
Introductions to other people.                     &      8 \\
\bottomrule
\end{tabular}
}

\texttt{\color{outcolor}Out[{\color{outcolor}111}]:}
    
    \centering{\begin{tabular}{lr}
\toprule
{} &  percent \\
\midrule
Writing code or otherwise implementing ideas.      &   56.45\% \\
Installing or using an application.                &   36.22\% \\
Other (please describe)                            &    3.56\% \\
Understanding community norms  &    3.49\% \\
Introductions to other people.                     &    0.28\% \\
\bottomrule
\end{tabular}
}

    \begin{center}
    \adjustimage{max size={0.9\linewidth}{0.9\paperheight}}{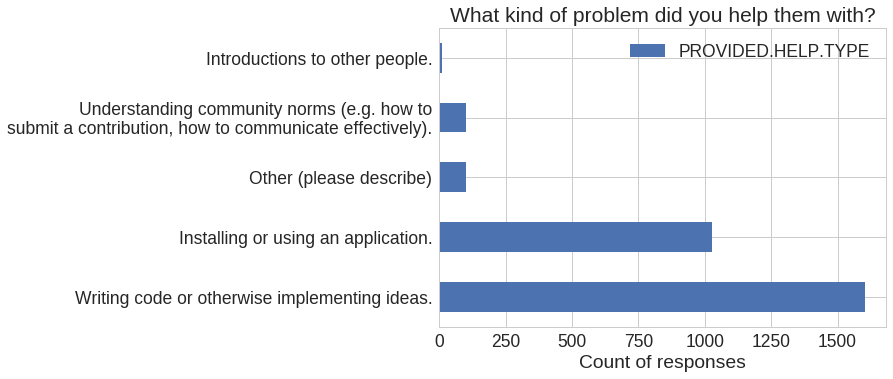}
    \end{center}
    { \hspace*{\fill} \\}
    \clearpage 
    \subsection{Open Source Software in Paid
Work}\label{open-source-software-in-paid-work}

      \subsubsection{Do you contribute to open source as part of your
professional
work?}\label{do-you-contribute-to-open-source-as-part-of-your-professional-work}

OSS.AS.JOB: Do you contribute to open source as part of your
professional work? In other words, are you paid for any of your time
spent on open source contributions?

\begin{itemize}
\tightlist
\item
  Yes, indirectly- I contribute to open source in carrying out my work
  duties, but I am not required or expected to do so.\\
\item
  No.\\
\item
  Yes, directly- some or all of my work duties include contributing to
  open source projects.
\end{itemize}

\texttt{\color{outcolor}Out[{\color{outcolor}113}]:}
    
    \centering{\begin{tabular}{lr}
\toprule
{} &  count \\
\midrule
Yes, indirectly &    896 \\
No.                                                &    687 \\
Yes, directly &    464 \\
\bottomrule
\end{tabular}
}

\texttt{\color{outcolor}Out[{\color{outcolor}114}]:}
    
    \centering{\begin{tabular}{lr}
\toprule
{} &  percent \\
\midrule
Yes, indirectly &   43.77\% \\
No.                                                &   33.56\% \\
Yes, directly &   22.67\% \\
\bottomrule
\end{tabular}
}

    \begin{center}
    \adjustimage{max size={0.9\linewidth}{0.9\paperheight}}{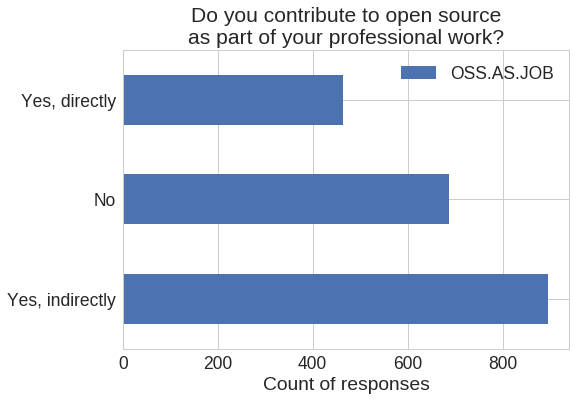}
    \end{center}
    { \hspace*{\fill} \\}
    
    \clearpage     \subsubsection{How often do you use open source software in your
professional
work?}\label{how-often-do-you-use-open-source-software-in-your-professional-work}

OSS.AT.WORK

\texttt{\color{outcolor}Out[{\color{outcolor}116}]:}
    
    \centering{\begin{tabular}{lr}
\toprule
{} &  count \\
\midrule
Frequently &   2191 \\
Sometimes  &    300 \\
Rarely     &    110 \\
Never      &     65 \\
\bottomrule
\end{tabular}
}

\texttt{\color{outcolor}Out[{\color{outcolor}117}]:}
    
    \centering{\begin{tabular}{lr}
\toprule
{} &  percent \\
\midrule
Frequently &   82.18\% \\
Sometimes  &   11.25\% \\
Rarely     &    4.13\% \\
Never      &    2.44\% \\
\bottomrule
\end{tabular}
}

    \begin{center}
    \adjustimage{max size={0.9\linewidth}{0.9\paperheight}}{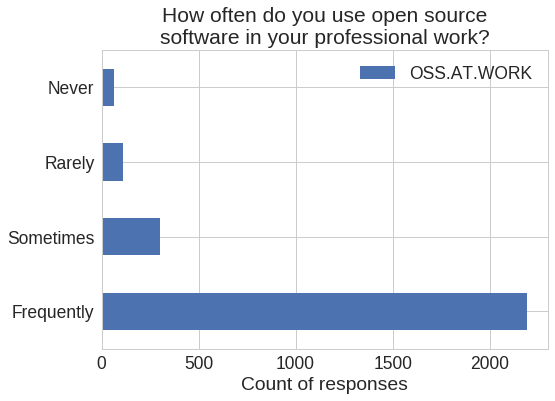}
    \end{center}
    { \hspace*{\fill} \\}
    
    \clearpage     \subsubsection{How does your employer's intellectual property
agreement/policy affect your free-time contributions to open source
unrelated to your
work?}\label{how-does-your-employers-intellectual-property-agreementpolicy-affect-your-free-time-contributions-to-open-source-unrelated-to-your-work}

OSS.IP.POLICY

\texttt{\color{outcolor}Out[{\color{outcolor}119}]:}
    
    \centering{\begin{tabular}{lr}
\toprule
{} &  count \\
\midrule
I am free to contribute without asking for permission. &           1178 \\
My employer doesn't have a clear policy on this.   &            695 \\
I am permitted to contribute to open source, but need to ask for permission &            287 \\
I'm not sure.                                      &            238 \\
Not applicable                                     &            180 \\
I am not permitted to contribute to open source at all. &             63 \\
\bottomrule
\end{tabular}
}

\texttt{\color{outcolor}Out[{\color{outcolor}120}]:}
    
    \centering{\begin{tabular}{lr}
\toprule
{} &  percent \\
\midrule
I am free to contribute without asking for permission. &   44.60\% \\
My employer doesn't have a clear policy on this.   &   26.32\% \\
I am permitted to contribute to open source, but need to ask for permission &   10.87\% \\
I'm not sure.                                      &    9.01\% \\
Not applicable                                     &    6.82\% \\
I am not permitted to contribute to open source at all. &    2.39\% \\
\bottomrule
\end{tabular}
}

    \begin{center}
    \adjustimage{max size={0.9\linewidth}{0.9\paperheight}}{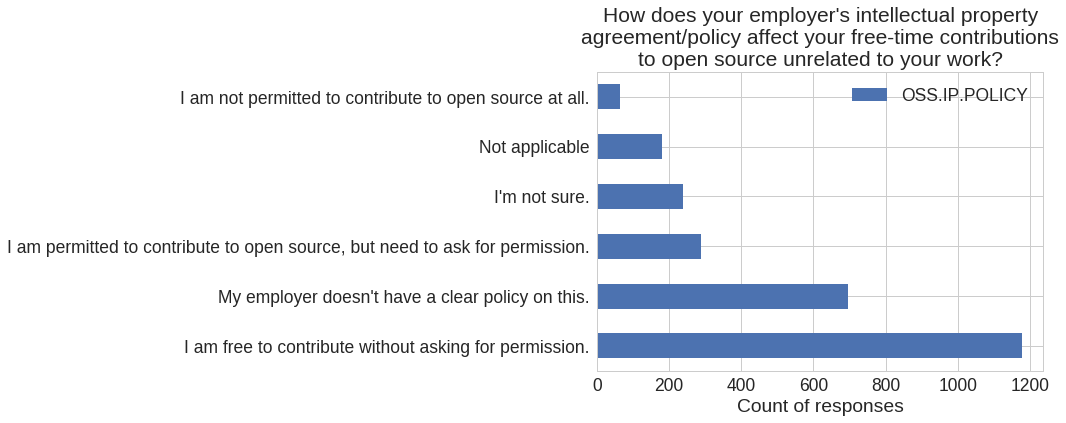}
    \end{center}
    { \hspace*{\fill} \\}
    
    \clearpage     \subsubsection{Which is closest to your employer's policy on using open
source software
applications?}\label{which-is-closest-to-your-employers-policy-on-using-open-source-software-applications}

\texttt{\color{outcolor}Out[{\color{outcolor}122}]:}
    
    \centering{\begin{tabular}{lr}
\toprule
{} &  count \\
\midrule
Use of open source applications is encouraged.     &                          1174 \\
Use of open source applications is acceptable if it is the most appropriate tool. &                           916 \\
My employer doesn't have a clear policy on this.   &                           338 \\
Not applicable                                     &                            88 \\
I'm not sure.                                      &                            83 \\
Use of open source applications is rarely, if ever, permitted. &                            42 \\
\bottomrule
\end{tabular}
}

\texttt{\color{outcolor}Out[{\color{outcolor}123}]:}
    
    \centering{\begin{tabular}{lr}
\toprule
{} &  percent \\
\midrule
Use of open source applications is encouraged.     &   44.45\% \\
Use of open source applications is acceptable if it is the most appropriate tool. &   34.68\% \\
My employer doesn't have a clear policy on this.   &   12.80\% \\
Not applicable                                     &    3.33\% \\
I'm not sure.                                      &    3.14\% \\
Use of open source applications is rarely, if ever, permitted. &    1.59\% \\
\bottomrule
\end{tabular}
}

    \begin{center}
    \adjustimage{max size={0.9\linewidth}{0.9\paperheight}}{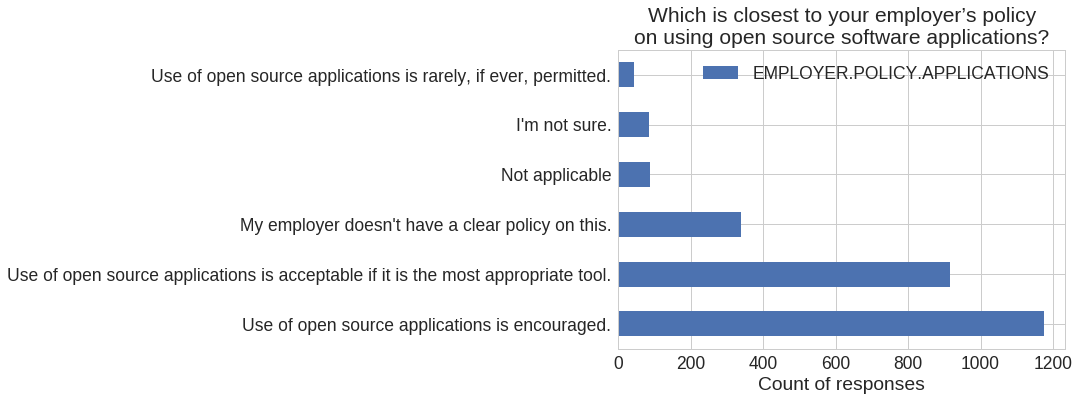}
    \end{center}
    { \hspace*{\fill} \\}
    
    \clearpage     \subsubsection{How important do you think your involvement in open
source was to getting your current
job?}\label{how-important-do-you-think-your-involvement-in-open-source-was-to-getting-your-current-job}

OSS.HIRING

\texttt{\color{outcolor}Out[{\color{outcolor}125}]:}
    
    \centering{\begin{tabular}{lr}
\toprule
{} &  count \\
\midrule
Very important                                     &    618 \\
Somewhat important                                 &    448 \\
Not at all important                               &    361 \\
Not too important                                  &    352 \\
Not applicable-I hadn't made any contributions &    254 \\
\bottomrule
\end{tabular}
}

\texttt{\color{outcolor}Out[{\color{outcolor}126}]:}
    
    \centering{\begin{tabular}{lr}
\toprule
{} &  percent \\
\midrule
Very important                                     &   30.40\% \\
Somewhat important                                 &   22.04\% \\
Not at all important                               &   17.76\% \\
Not too important                                  &   17.31\% \\
Not applicable-I hadn't made any contributions &   12.49\% \\
\bottomrule
\end{tabular}
}

    \begin{center}
    \adjustimage{max size={0.9\linewidth}{0.9\paperheight}}{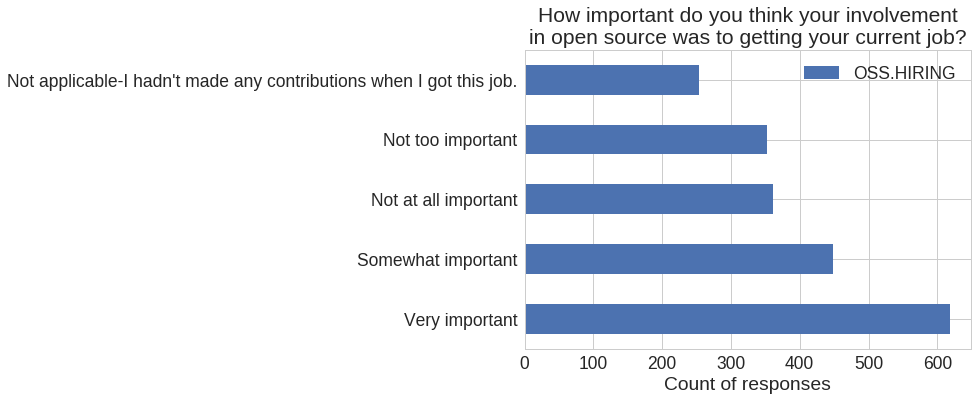}
    \end{center}
    { \hspace*{\fill} \\}
    \clearpage
    \subsection{Demographics}\label{demographics}

         \subsubsection{Do you currently live in a country other than the one in
which you were
born?}\label{do-you-currently-live-in-a-country-other-than-the-one-in-which-you-were-born}

IMMIGRATION

\texttt{\color{outcolor}Out[{\color{outcolor}128}]:}
    
    \centering{\begin{tabular}{lr}
\toprule
{} &  count \\
\midrule
No, I live in the country where I was born.   &   2764 \\
Yes, and I intend to stay permanently.        &    513 \\
Yes, and I am not sure about my future plans. &    292 \\
Yes, and I intend to stay temporarily.        &    165 \\
\bottomrule
\end{tabular}
}

\texttt{\color{outcolor}Out[{\color{outcolor}129}]:}
    
    \centering{\begin{tabular}{lr}
\toprule
{} &  percent \\
\midrule
No, I live in the country where I was born.   &   74.02\% \\
Yes, and I intend to stay permanently.        &   13.74\% \\
Yes, and I am not sure about my future plans. &    7.82\% \\
Yes, and I intend to stay temporarily.        &    4.42\% \\
\bottomrule
\end{tabular}
}

    \begin{center}
    \adjustimage{max size={0.9\linewidth}{0.9\paperheight}}{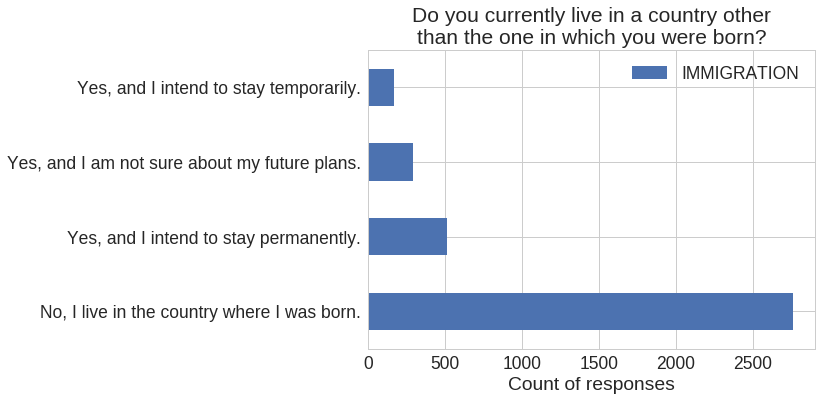}
    \end{center}
    { \hspace*{\fill} \\}
    
    \clearpage     \subsubsection{Thinking of where you were born, are you a member of an
ethnicity or nationality that is a considered a minority in that
country?}\label{thinking-of-where-you-were-born-are-you-a-member-of-an-ethnicity-or-nationality-that-is-a-considered-a-minority-in-that-country}

MINORITY.HOMECOUNTRY

\texttt{\color{outcolor}Out[{\color{outcolor}131}]:}
    
    \centering{\begin{tabular}{lr}
\toprule
{} &  count \\
\midrule
No                &    754 \\
Yes               &    124 \\
Not sure          &     45 \\
Prefer not to say &     34 \\
\bottomrule
\end{tabular}
}

\texttt{\color{outcolor}Out[{\color{outcolor}132}]:}
    
    \centering{\begin{tabular}{lr}
\toprule
{} &  percent \\
\midrule
No                &   78.79\% \\
Yes               &   12.96\% \\
Not sure          &    4.70\% \\
Prefer not to say &    3.55\% \\
\bottomrule
\end{tabular}
}

    \begin{center}
    \adjustimage{max size={0.9\linewidth}{0.9\paperheight}}{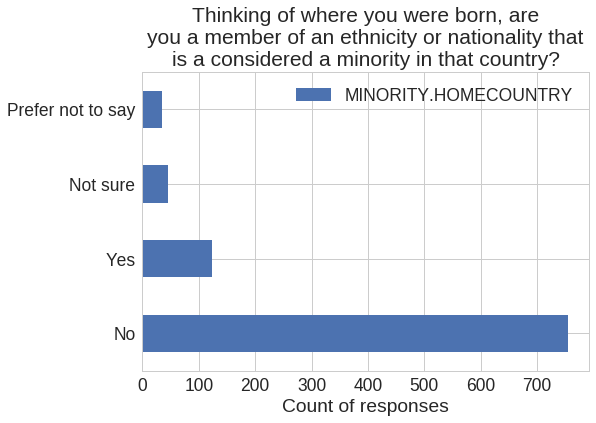}
    \end{center}
    { \hspace*{\fill} \\}

    \clearpage     \subsubsection{Thinking of where you currently live, are you a member of
an ethnicity or nationality that is a considered a minority in that
country?}\label{thinking-of-where-you-currently-live-are-you-a-member-of-an-ethnicity-or-nationality-that-is-a-considered-a-minority-in-that-country}

MINORITY.CURRENT.COUNTRY

\texttt{\color{outcolor}Out[{\color{outcolor}134}]:}
    
    \centering{\begin{tabular}{lr}
\toprule
{} &  count \\
\midrule
No                &   2837 \\
Yes               &    546 \\
Not sure          &    193 \\
Prefer not to say &    156 \\
\bottomrule
\end{tabular}
}

\texttt{\color{outcolor}Out[{\color{outcolor}135}]:}
    
    \centering{\begin{tabular}{lr}
\toprule
{} &  percent \\
\midrule
No                &   76.02\% \\
Yes               &   14.63\% \\
Not sure          &    5.17\% \\
Prefer not to say &    4.18\% \\
\bottomrule
\end{tabular}
}

    \begin{center}
    \adjustimage{max size={0.9\linewidth}{0.9\paperheight}}{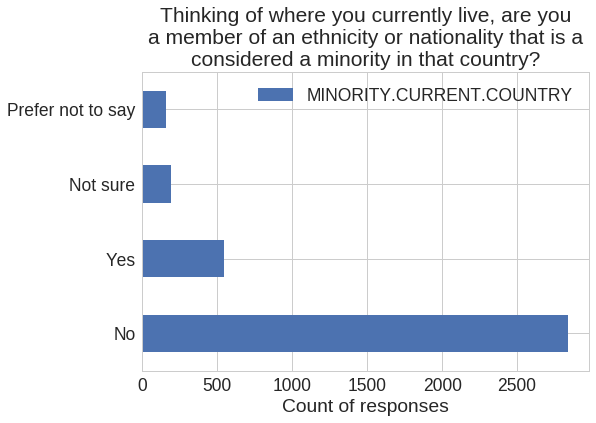}
    \end{center}
    { \hspace*{\fill} \\}
    
    \clearpage     \subsubsection{What is your gender?}\label{what-is-your-gender}

GENDER

\texttt{\color{outcolor}Out[{\color{outcolor}137}]:}
    
    \centering{\begin{tabular}{lr}
\toprule
{} &  count \\
\midrule
Man                  &   3387 \\
Prefer not to say    &    173 \\
Woman                &    125 \\
Non-binary  or Other &     39 \\
\bottomrule
\end{tabular}
}

\texttt{\color{outcolor}Out[{\color{outcolor}138}]:}
    
    \centering{\begin{tabular}{lr}
\toprule
{} &  percent \\
\midrule
Man                  &   90.95\% \\
Prefer not to say    &    4.65\% \\
Woman                &    3.36\% \\
Non-binary  or Other &    1.05\% \\
\bottomrule
\end{tabular}
}

    \begin{center}
    \adjustimage{max size={0.9\linewidth}{0.9\paperheight}}{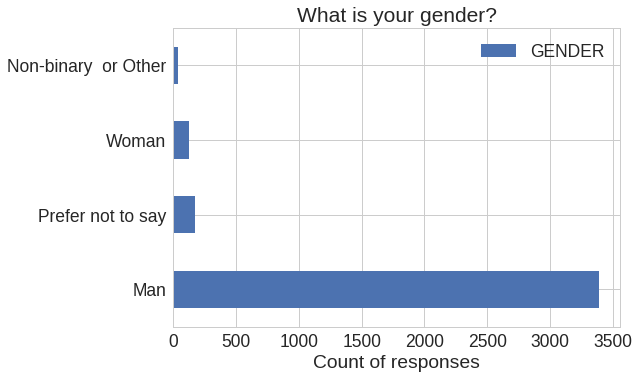}
    \end{center}
    { \hspace*{\fill} \\}
    
    \clearpage     \subsubsection{Do you identify as
transgender?}\label{do-you-identify-as-transgender}

TRANSGENDER.IDENTITY

\texttt{\color{outcolor}Out[{\color{outcolor}140}]:}
    
    \centering{\begin{tabular}{lr}
\toprule
{} &  count \\
\midrule
No                &   3494 \\
Prefer not to say &    158 \\
Yes               &     33 \\
Not sure          &     30 \\
\bottomrule
\end{tabular}
}

\texttt{\color{outcolor}Out[{\color{outcolor}141}]:}
    
    \centering{\begin{tabular}{lr}
\toprule
{} &  percent \\
\midrule
No                &   94.05\% \\
Prefer not to say &    4.25\% \\
Yes               &    0.89\% \\
Not sure          &    0.81\% \\
\bottomrule
\end{tabular}
}

    \begin{center}
    \adjustimage{max size={0.9\linewidth}{0.9\paperheight}}{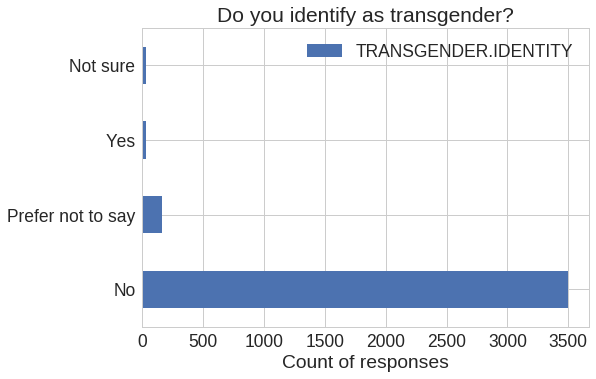}
    \end{center}
    { \hspace*{\fill} \\}

    \clearpage     \subsubsection{Do you identify as gay, lesbian, or bisexual, asexual, or
any other minority sexual
orientation?}\label{do-you-identify-as-gay-lesbian-or-bisexual-asexual-or-any-other-minority-sexual-orientation}

SEXUAL.ORIENTATION

\texttt{\color{outcolor}Out[{\color{outcolor}143}]:}
    
    \centering{\begin{tabular}{lr}
\toprule
{} &  count \\
\midrule
No                &   3187 \\
Yes               &    246 \\
Prefer not to say &    201 \\
Not sure          &     85 \\
\bottomrule
\end{tabular}
}

\texttt{\color{outcolor}Out[{\color{outcolor}144}]:}
    
    \centering{\begin{tabular}{lr}
\toprule
{} &  percent \\
\midrule
No                &   85.70\% \\
Yes               &    6.61\% \\
Prefer not to say &    5.40\% \\
Not sure          &    2.29\% \\
\bottomrule
\end{tabular}
}

    \begin{center}
    \adjustimage{max size={0.9\linewidth}{0.9\paperheight}}{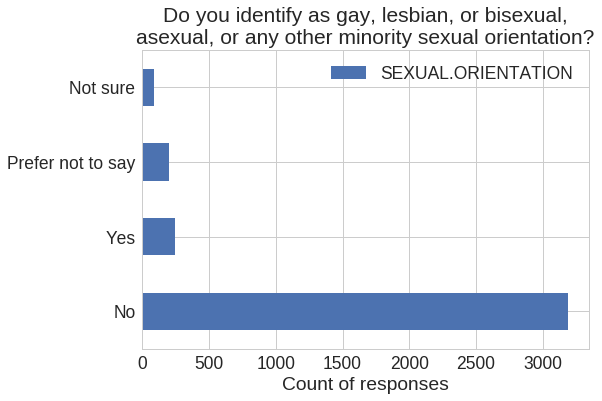}
    \end{center}
    { \hspace*{\fill} \\}
    
    \clearpage     \subsubsection{How well can you read and write in
English?}\label{how-well-can-you-read-and-write-in-english}

WRITTEN.ENGLISH

\texttt{\color{outcolor}Out[{\color{outcolor}146}]:}
    
    \centering{\begin{tabular}{lr}
\toprule
{} &  count \\
\midrule
Very well       &   2865 \\
Moderately well &    742 \\
Not very well   &    108 \\
Not at all      &      6 \\
\bottomrule
\end{tabular}
}

\texttt{\color{outcolor}Out[{\color{outcolor}147}]:}
    
    \centering{\begin{tabular}{lr}
\toprule
{} &  percent \\
\midrule
Very well       &   77.00\% \\
Moderately well &   19.94\% \\
Not very well   &    2.90\% \\
Not at all      &    0.16\% \\
\bottomrule
\end{tabular}
}

    \begin{center}
    \adjustimage{max size={0.9\linewidth}{0.9\paperheight}}{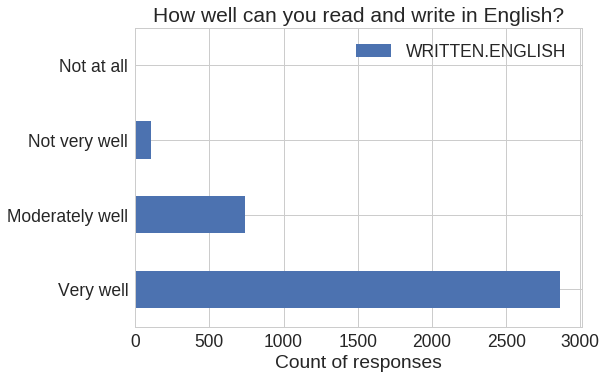}
    \end{center}
    { \hspace*{\fill} \\}

    \clearpage     \subsubsection{What is your age?}\label{what-is-your-age}

AGE

\texttt{\color{outcolor}Out[{\color{outcolor}149}]:}
    
    \centering{\begin{tabular}{lr}
\toprule
{} &  count \\
\midrule
17 or younger     &    139 \\
18 to 24 years    &    871 \\
25 to 34 years    &   1400 \\
35 to 44 years    &    772 \\
45 to 54 years    &    267 \\
55 to 64 years    &     93 \\
65 years or older &     36 \\
\bottomrule
\end{tabular}
}

\texttt{\color{outcolor}Out[{\color{outcolor}150}]:}
    
    \centering{\begin{tabular}{lr}
\toprule
{} &  percent \\
\midrule
17 or younger     &    3.88\% \\
18 to 24 years    &   24.34\% \\
25 to 34 years    &   39.13\% \\
35 to 44 years    &   21.58\% \\
45 to 54 years    &    7.46\% \\
55 to 64 years    &    2.60\% \\
65 years or older &    1.01\% \\
\bottomrule
\end{tabular}
}

    \begin{center}
    \adjustimage{max size={0.9\linewidth}{0.9\paperheight}}{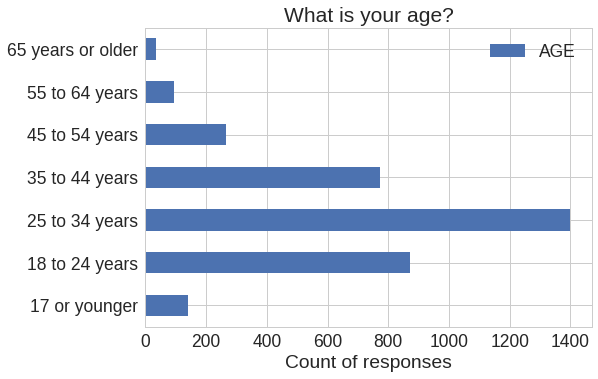}
    \end{center}
    { \hspace*{\fill} \\}

    \clearpage     \subsubsection{What is highest level of formal education that you have
completed?}\label{what-is-highest-level-of-formal-education-that-you-have-completed}

FORMAL.EDUCATION

\texttt{\color{outcolor}Out[{\color{outcolor}152}]:}
    
    \centering{\begin{tabular}{lr}
\toprule
{} &  count \\
\midrule
Bachelor's degree                                  &   1321 \\
Master's degree                                    &    852 \\
Some college, no degree                            &    640 \\
Secondary (high) school graduate or equivalent     &    375 \\
Doctorate (Ph.D.) or other advanced degree  &    256 \\
Vocational/trade program or apprenticeship         &    127 \\
Less than secondary (high) school                  &    126 \\
\bottomrule
\end{tabular}
}

\texttt{\color{outcolor}Out[{\color{outcolor}153}]:}
    
    \centering{\begin{tabular}{lr}
\toprule
{} &  percent \\
\midrule
Bachelor's degree                                  &   35.73\% \\
Master's degree                                    &   23.05\% \\
Some college, no degree                            &   17.31\% \\
Secondary (high) school graduate or equivalent     &   10.14\% \\
Doctorate (Ph.D.) or other advanced degree  &    6.92\% \\
Vocational/trade program or apprenticeship         &    3.44\% \\
Less than secondary (high) school                  &    3.41\% \\
\bottomrule
\end{tabular}
}

    \begin{center}
    \adjustimage{max size={0.9\linewidth}{0.9\paperheight}}{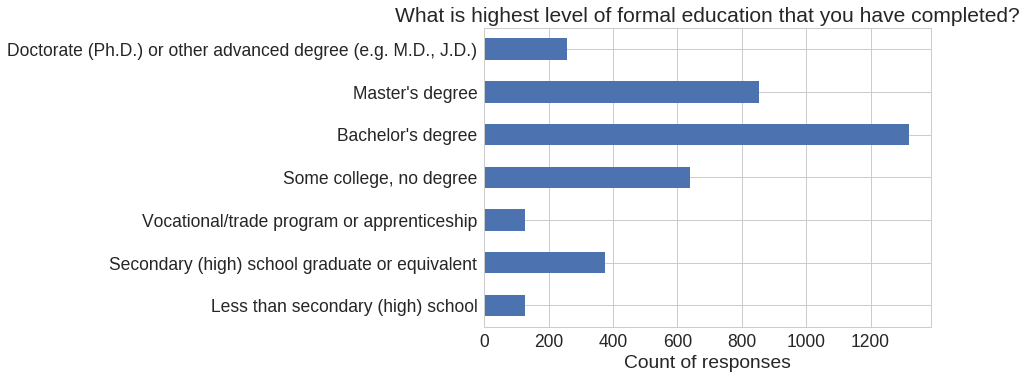}
    \end{center}
    { \hspace*{\fill} \\}

    \clearpage     \subsubsection{What is the highest level of formal education that either
of your parents
completed?}\label{what-is-the-highest-level-of-formal-education-that-either-of-your-parents-completed}

PARENTS.FORMAL.EDUCATION

\texttt{\color{outcolor}Out[{\color{outcolor}155}]:}
    
    \centering{\begin{tabular}{lr}
\toprule
{} &  count \\
\midrule
Bachelor's degree                                  &    961 \\
Master's degree                                    &    871 \\
Secondary (high) school graduate or equivalent     &    566 \\
Some college, no degree                            &    388 \\
Doctorate (Ph.D.) or other advanced degree  &    387 \\
Vocational/trade program or apprenticeship         &    257 \\
Less than secondary (high) school                  &    243 \\
\bottomrule
\end{tabular}
}

\texttt{\color{outcolor}Out[{\color{outcolor}156}]:}
    
    \centering{\begin{tabular}{lr}
\toprule
{} &  percent \\
\midrule
Bachelor's degree                                  &   26.16\% \\
Master's degree                                    &   23.71\% \\
Secondary (high) school graduate or equivalent     &   15.41\% \\
Some college, no degree                            &   10.56\% \\
Doctorate (Ph.D.) or other advanced degree  &   10.54\% \\
Vocational/trade program or apprenticeship         &    7.00\% \\
Less than secondary (high) school                  &    6.62\% \\
\bottomrule
\end{tabular}
}

    \begin{center}
    \adjustimage{max size={0.9\linewidth}{0.9\paperheight}}{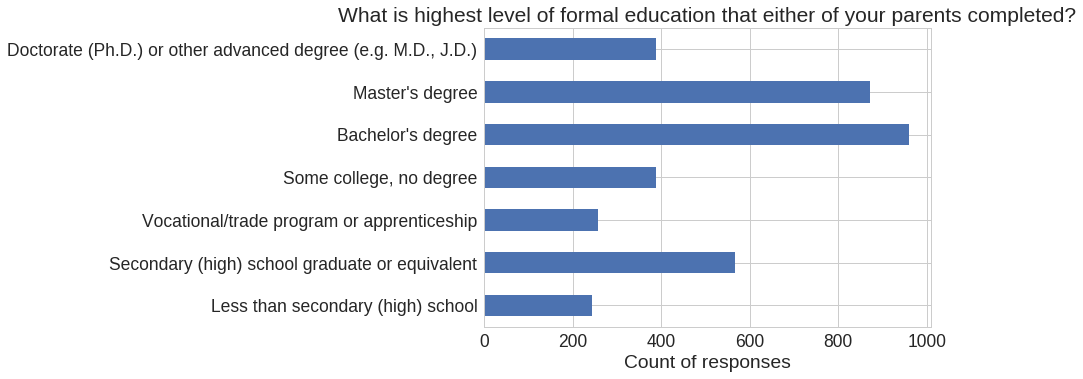}
    \end{center}
    { \hspace*{\fill} \\}

    \clearpage     \subsubsection{How old were you when you first had regular access to a
computer with an internet
connection?}\label{how-old-were-you-when-you-first-had-regular-access-to-a-computer-with-an-internet-connection}

AGE.AT.FIRST.COMPUTER.INTERNET

\texttt{\color{outcolor}Out[{\color{outcolor}158}]:}
    
    \centering{\begin{tabular}{lr}
\toprule
{} &  count \\
\midrule
Younger than 13 years old &   1478 \\
13 - 17 years old         &   1313 \\
18 - 24 years old         &    695 \\
25 - 45 years old         &    202 \\
Older than 45 years old   &     23 \\
\bottomrule
\end{tabular}
}

\texttt{\color{outcolor}Out[{\color{outcolor}159}]:}
    
    \centering{\begin{tabular}{lr}
\toprule
{} &  percent \\
\midrule
Younger than 13 years old &   39.83\% \\
13 - 17 years old         &   35.38\% \\
18 - 24 years old         &   18.73\% \\
25 - 45 years old         &    5.44\% \\
Older than 45 years old   &    0.62\% \\
\bottomrule
\end{tabular}
}

    \begin{center}
    \adjustimage{max size={0.9\linewidth}{0.9\paperheight}}{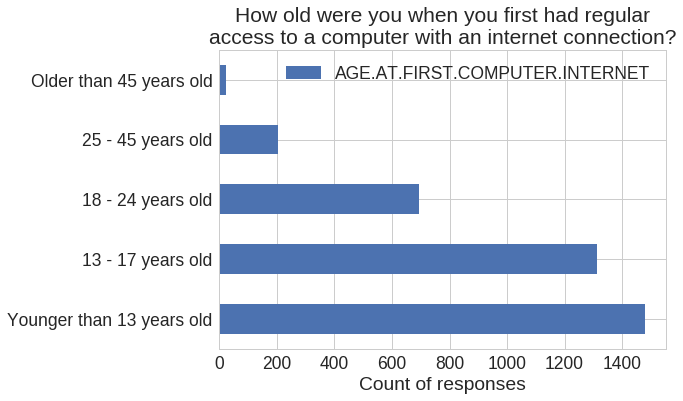}
    \end{center}
    { \hspace*{\fill} \\}

    \clearpage     \subsubsection{Where did you first have regular access to a computer
with internet
connection?}\label{where-did-you-first-have-regular-access-to-a-computer-with-internet-connection}

LOCATION.OF.FIRST.COMPUTER.INTERNET

\texttt{\color{outcolor}Out[{\color{outcolor}161}]:}
    
    \centering{\begin{tabular}{lr}
\toprule
{} &  count \\
\midrule
At home (belonging to me or a family member)       &   2520 \\
In a classroom, computer lab, or library at school &    746 \\
At an internet cafe or similar space               &    182 \\
Other (please describe)                            &    106 \\
At a public library or community center            &     87 \\
At work (recoded from open ended)                  &     70 \\
\bottomrule
\end{tabular}
}

\texttt{\color{outcolor}Out[{\color{outcolor}162}]:}
    
    \centering{\begin{tabular}{lr}
\toprule
{} &  percent \\
\midrule
At home (belonging to me or a family member)       &   67.91\% \\
In a classroom, computer lab, or library at school &   20.10\% \\
At an internet cafe or similar space               &    4.90\% \\
Other (please describe)                            &    2.86\% \\
At a public library or community center            &    2.34\% \\
At work (recoded from open ended)                  &    1.89\% \\
\bottomrule
\end{tabular}
}

    \begin{center}
    \adjustimage{max size={0.9\linewidth}{0.9\paperheight}}{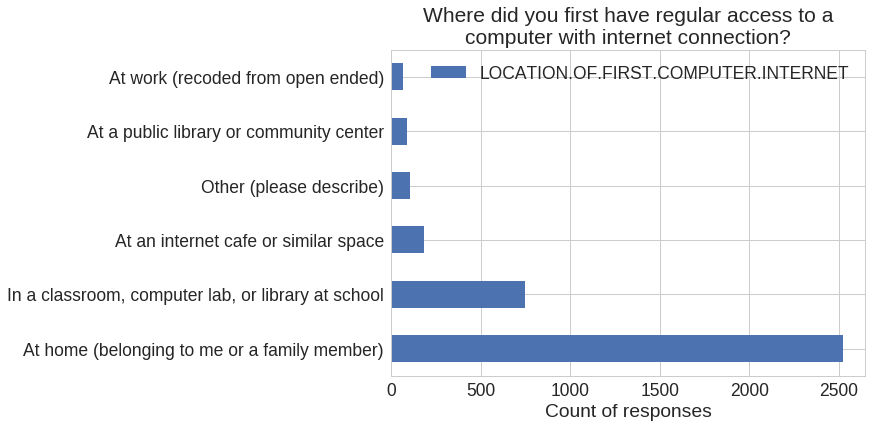}
    \end{center}
    { \hspace*{\fill} \\}
    
    \clearpage     \subsubsection{Where was the respondent surveyed
from?}\label{where-was-the-respondent-surveyed-from}

POPLATION

\texttt{\color{outcolor}Out[{\color{outcolor}164}]:}
    
    \centering{\begin{tabular}{lr}
\toprule
{} &  count \\
\midrule
github             &   5495 \\
off site community &    534 \\
\bottomrule
\end{tabular}
}

\texttt{\color{outcolor}Out[{\color{outcolor}165}]:}
    
    \centering{\begin{tabular}{lr}
\toprule
{} &  percent \\
\midrule
github             &   91.14\% \\
off site community &    8.86\% \\
\bottomrule
\end{tabular}
}

    \begin{center}
    \adjustimage{max size={0.9\linewidth}{0.9\paperheight}}{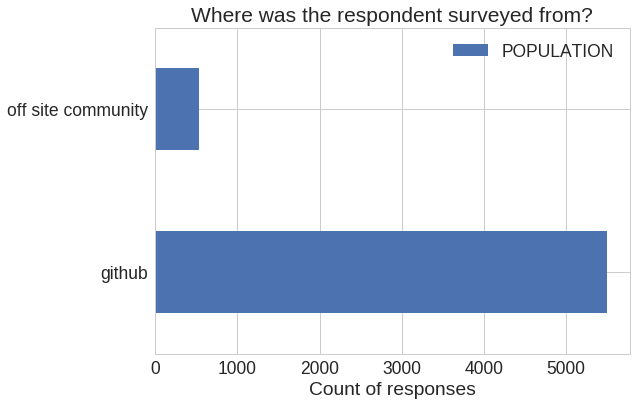}
    \end{center}
    { \hspace*{\fill} \\}
    \clearpage 
    \subsection{Harassment / Inclusiveness of
OSS}\label{harassment-inclusiveness-of-oss}

        \subsubsection{Have you ever observed any of the following in the
context of an open source
project?}\label{have-you-ever-observed-any-of-the-following-in-the-context-of-an-open-source-project}

DISCOURAGING.BEHAVIOR.*

\texttt{\color{outcolor}Out[{\color{outcolor}167}]:}
    
    \centering{\begin{tabular}{lrr}
\toprule
{} &   Yes &    No \\
\midrule
DISCOURAGING.BEHAVIOR.LACK.OF.RESPONSE           &  3017 &   792 \\
DISCOURAGING.BEHAVIOR.REJECTION.WOUT.EXPLANATION &  1210 &  2580 \\
DISCOURAGING.BEHAVIOR.DISMISSIVE.RESPONSE        &  2195 &  1598 \\
DISCOURAGING.BEHAVIOR.BAD.DOCS                   &  3559 &   263 \\
DISCOURAGING.BEHAVIOR.CONFLICT                   &  1830 &  1966 \\
DISCOURAGING.BEHAVIOR.UNWELCOMING.LANGUAGE       &   649 &  3158 \\
\bottomrule
\end{tabular}
}

\texttt{\color{outcolor}Out[{\color{outcolor}168}]:}
    
    \centering{\begin{tabular}{lr}
\toprule
{} &  percent\_yes \\
\midrule
DISCOURAGING.BEHAVIOR.UNWELCOMING.LANGUAGE       &       17.05\% \\
DISCOURAGING.BEHAVIOR.REJECTION.WOUT.EXPLANATION &       31.93\% \\
DISCOURAGING.BEHAVIOR.CONFLICT                   &       48.21\% \\
DISCOURAGING.BEHAVIOR.DISMISSIVE.RESPONSE        &       57.87\% \\
DISCOURAGING.BEHAVIOR.LACK.OF.RESPONSE           &       79.21\% \\
DISCOURAGING.BEHAVIOR.BAD.DOCS                   &       93.12\% \\
\bottomrule
\end{tabular}
}

    \begin{center}
    \adjustimage{max size={0.9\linewidth}{0.9\paperheight}}{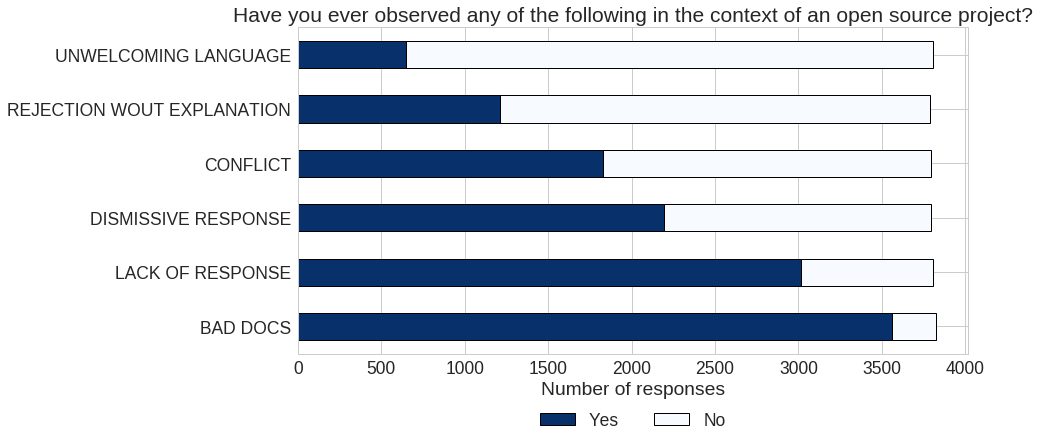}
    \end{center}
    { \hspace*{\fill} \\}

    \clearpage     \subsubsection{Have you ever witnessed any of the following behaviors
directed at another person in the context of an open source project?
(not including something directed at
you)}\label{have-you-ever-witnessed-any-of-the-following-behaviors-directed-at-another-person-in-the-context-of-an-open-source-project-not-including-something-directed-at-you}

NEGATIVE.WITNESS.*

\texttt{\color{outcolor}Out[{\color{outcolor}172}]:}
    
    \centering{\begin{tabular}{lrr}
\toprule
{} &   Yes &    No \\
\midrule
NEGATIVE.WITNESS.RUDENESS                  &  1753 &  1911 \\
NEGATIVE.WITNESS.NAME.CALLING              &   789 &  2875 \\
NEGATIVE.WITNESS.THREATS                   &   162 &  3502 \\
NEGATIVE.WITNESS.IMPERSONATION             &   177 &  3487 \\
NEGATIVE.WITNESS.SUSTAINED.HARASSMENT      &   237 &  3427 \\
NEGATIVE.WITNESS.CROSS.PLATFORM.HARASSMENT &   175 &  3489 \\
NEGATIVE.WITNESS.STALKING                  &   108 &  3556 \\
NEGATIVE.WITNESS.SEXUAL.ADVANCES           &   136 &  3528 \\
NEGATIVE.WITNESS.STEREOTYPING              &   423 &  3241 \\
NEGATIVE.WITNESS.DOXXING                   &   151 &  3513 \\
NEGATIVE.WITNESS.OTHER                     &    78 &  3586 \\
NEGATIVE.WITNESS.NONE.OF.THE.ABOVE         &  1721 &  1943 \\
\bottomrule
\end{tabular}
}

\texttt{\color{outcolor}Out[{\color{outcolor}173}]:}
    
    \centering{\begin{tabular}{lr}
\toprule
{} &  percent\_yes \\
\midrule
NEGATIVE.WITNESS.OTHER                     &        2.13\% \\
NEGATIVE.WITNESS.STALKING                  &        2.95\% \\
NEGATIVE.WITNESS.SEXUAL.ADVANCES           &        3.71\% \\
NEGATIVE.WITNESS.DOXXING                   &        4.12\% \\
NEGATIVE.WITNESS.THREATS                   &        4.42\% \\
NEGATIVE.WITNESS.CROSS.PLATFORM.HARASSMENT &        4.78\% \\
NEGATIVE.WITNESS.IMPERSONATION             &        4.83\% \\
NEGATIVE.WITNESS.SUSTAINED.HARASSMENT      &        6.47\% \\
NEGATIVE.WITNESS.STEREOTYPING              &       11.54\% \\
NEGATIVE.WITNESS.NAME.CALLING              &       21.53\% \\
NEGATIVE.WITNESS.NONE.OF.THE.ABOVE         &       46.97\% \\
NEGATIVE.WITNESS.RUDENESS                  &       47.84\% \\
\bottomrule
\end{tabular}
}

    \begin{center}
    \adjustimage{max size={0.85\linewidth}{0.9\paperheight}}{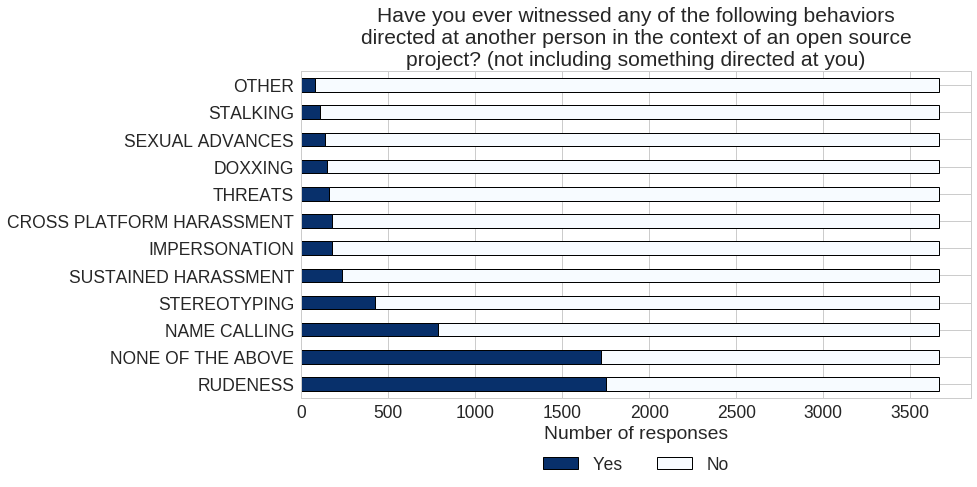}
    \end{center}
    { \hspace*{\fill} \\}
    
    \subsubsection{Have you ever experienced any of the following behaviors
directed at you in the context of an open source
project?}\label{have-you-ever-experienced-any-of-the-following-behaviors-directed-at-you-in-the-context-of-an-open-source-project}

\texttt{\color{outcolor}Out[{\color{outcolor}177}]:}
    
    \centering{\begin{tabular}{lrr}
\toprule
{} &   Yes &    No \\
\midrule
NEGATIVE.EXPERIENCE.RUDENESS                  &   646 &  2992 \\
NEGATIVE.EXPERIENCE.NAME.CALLING              &   192 &  3446 \\
NEGATIVE.EXPERIENCE.THREATS                   &    43 &  3595 \\
NEGATIVE.EXPERIENCE.IMPERSONATION             &    45 &  3593 \\
NEGATIVE.EXPERIENCE.SUSTAINED.HARASSMENT      &    55 &  3583 \\
NEGATIVE.EXPERIENCE.CROSS.PLATFORM.HARASSMENT &    42 &  3596 \\
NEGATIVE.EXPERIENCE.STALKING                  &    35 &  3603 \\
NEGATIVE.EXPERIENCE.SEXUAL.ADVANCES           &    25 &  3613 \\
NEGATIVE.EXPERIENCE.STEREOTYPING              &   114 &  3524 \\
NEGATIVE.EXPERIENCE.DOXXING                   &    23 &  3615 \\
NEGATIVE.EXPERIENCE.OTHER                     &    39 &  3599 \\
NEGATIVE.EXPERIENCE.NONE.OF.THE.ABOVE         &  2900 &   738 \\
\bottomrule
\end{tabular}
}

\texttt{\color{outcolor}Out[{\color{outcolor}178}]:}
    
    \centering{\begin{tabular}{lr}
\toprule
{} &  percent\_yes \\
\midrule
NEGATIVE.EXPERIENCE.DOXXING                   &        0.63\% \\
NEGATIVE.EXPERIENCE.SEXUAL.ADVANCES           &        0.69\% \\
NEGATIVE.EXPERIENCE.STALKING                  &        0.96\% \\
NEGATIVE.EXPERIENCE.OTHER                     &        1.07\% \\
NEGATIVE.EXPERIENCE.CROSS.PLATFORM.HARASSMENT &        1.15\% \\
NEGATIVE.EXPERIENCE.THREATS                   &        1.18\% \\
NEGATIVE.EXPERIENCE.IMPERSONATION             &        1.24\% \\
NEGATIVE.EXPERIENCE.SUSTAINED.HARASSMENT      &        1.51\% \\
NEGATIVE.EXPERIENCE.STEREOTYPING              &        3.13\% \\
NEGATIVE.EXPERIENCE.NAME.CALLING              &        5.28\% \\
NEGATIVE.EXPERIENCE.RUDENESS                  &       17.76\% \\
NEGATIVE.EXPERIENCE.NONE.OF.THE.ABOVE         &       79.71\% \\
\bottomrule
\end{tabular}
}

    \begin{center}
    \adjustimage{max size={0.85\linewidth}{0.9\paperheight}}{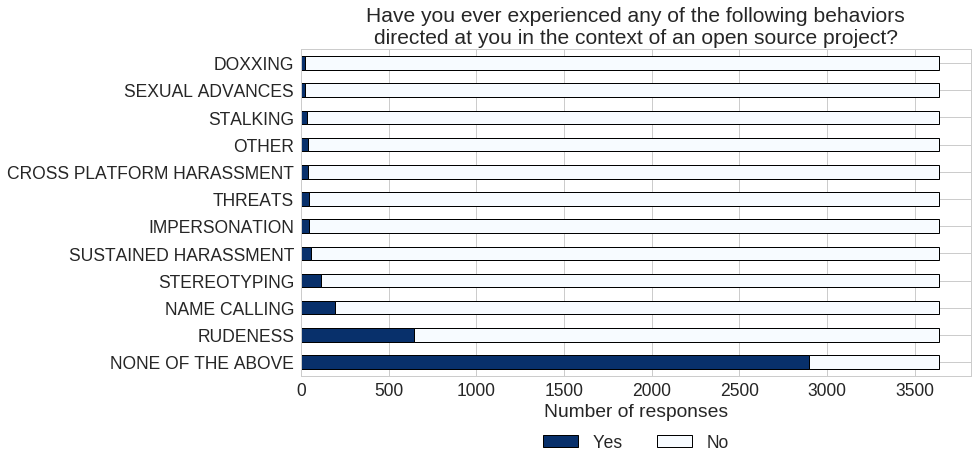}
    \end{center}
    { \hspace*{\fill} \\}
    
    \subsubsection{Thinking of the last time you experienced harassment, how
did you
respond?}\label{thinking-of-the-last-time-you-experienced-harassment-how-did-you-respond}

NEGATIVE.RESPONSE.*

\texttt{\color{outcolor}Out[{\color{outcolor}182}]:}
    
    \centering{\begin{tabular}{lrr}
\toprule
{} &  Yes &   No \\
\midrule
NEGATIVE.RESPONSE.ASKED.USER.TO.STOP          &  194 &  525 \\
NEGATIVE.RESPONSE.SOLICITED.COMMUNITY.SUPPORT &  112 &  607 \\
NEGATIVE.RESPONSE.BLOCKED.USER                &  170 &  549 \\
NEGATIVE.RESPONSE.REPORTED.TO.MAINTAINERS     &   95 &  624 \\
NEGATIVE.RESPONSE.REPORTED.TO.HOST.OR.ISP     &   20 &  699 \\
NEGATIVE.RESPONSE.CONSULTED.LEGAL.COUNSEL     &    8 &  711 \\
NEGATIVE.RESPONSE.CONTACTED.LAW.ENFORCEMENT   &    9 &  710 \\
NEGATIVE.RESPONSE.OTHER                       &   71 &  648 \\
NEGATIVE.RESPONSE.IGNORED                     &  350 &  369 \\
\bottomrule
\end{tabular}
}

\texttt{\color{outcolor}Out[{\color{outcolor}183}]:}
    
    \centering{\begin{tabular}{lr}
\toprule
{} &  percent\_yes \\
\midrule
NEGATIVE.RESPONSE.CONSULTED.LEGAL.COUNSEL     &        1.11\% \\
NEGATIVE.RESPONSE.CONTACTED.LAW.ENFORCEMENT   &        1.25\% \\
NEGATIVE.RESPONSE.REPORTED.TO.HOST.OR.ISP     &        2.78\% \\
NEGATIVE.RESPONSE.OTHER                       &        9.87\% \\
NEGATIVE.RESPONSE.REPORTED.TO.MAINTAINERS     &       13.21\% \\
NEGATIVE.RESPONSE.SOLICITED.COMMUNITY.SUPPORT &       15.58\% \\
NEGATIVE.RESPONSE.BLOCKED.USER                &       23.64\% \\
NEGATIVE.RESPONSE.ASKED.USER.TO.STOP          &       26.98\% \\
NEGATIVE.RESPONSE.IGNORED                     &       48.68\% \\
\bottomrule
\end{tabular}
}

    \begin{center}
    \adjustimage{max size={0.85\linewidth}{0.9\paperheight}}{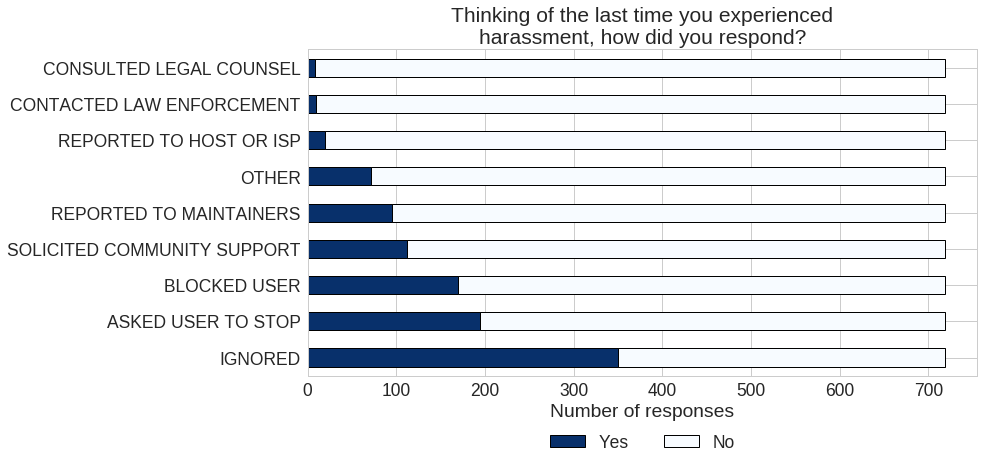}
    \end{center}
    { \hspace*{\fill} \\}

    \clearpage     \subsubsection{How effective were the following
responses? Response counts}\label{how-effective-were-the-following-responses-counts}

RESPONSE.EFFECTIVENESS.*

\texttt{\color{outcolor}Out[{\color{outcolor}186}]:}
    
    \centering{\begin{tabular}{lrrrrr}
\toprule
{} &  \pbox{10cm}{Not at all\\ effective} &  \pbox{10cm}{A little\\ effective} &  \pbox{10cm}{Somewhat\\ effective} &  \pbox{10cm}{Mostly\\ effective} &  \pbox{10cm}{Completely\\ effective} \\
\midrule
CONTACTED LAW ENFORCEMENT   &                 4 &               0 &               2 &             0 &                 3 \\
CONSULTED LEGAL COUNSEL     &                 1 &               1 &               3 &             2 &                 1 \\
REPORTED TO HOST OR ISP     &                 6 &               4 &               6 &             3 &                 1 \\
OTHER                       &                 4 &               0 &               4 &            10 &                11 \\
REPORTED TO MAINTAINERS     &                10 &              11 &              31 &            30 &                13 \\
SOLICITED COMMUNITY SUPPORT &                 6 &              22 &              38 &            32 &                14 \\
ASKED USER TO STOP          &                48 &              51 &              50 &            33 &                11 \\
BLOCKED USER                &                 6 &              20 &              28 &            56 &                58 \\
\bottomrule
\end{tabular}
}

    \begin{center}
    \adjustimage{max size={1\linewidth}{0.9\paperheight}}{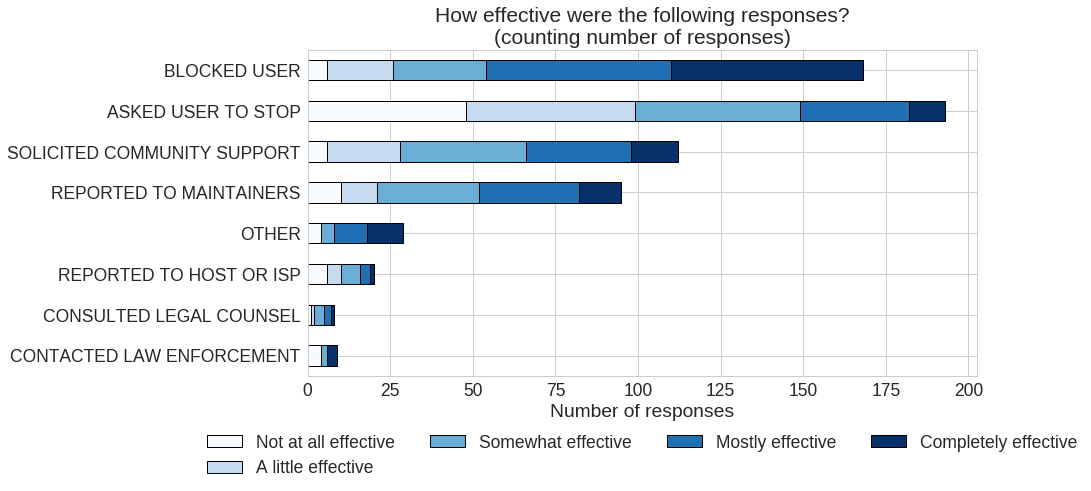}
    \end{center}
    { \hspace*{\fill} \\}
    
    \clearpage     \subsubsection{How effective were the following
responses? Proportions}\label{how-effective-were-the-following-responses-prop}
\texttt{\color{outcolor}Out[{\color{outcolor}189}]:}
    
    \centering{\begin{tabular}{lrrrrr}
\toprule
{} &  \pbox{10cm}{Not at all\\ effective} &  \pbox{10cm}{A little\\ effective} &  \pbox{10cm}{Somewhat\\ effective} &  \pbox{10cm}{Mostly\\ effective} &  \pbox{10cm}{Completely\\ effective} \\
\midrule
REPORTED TO HOST OR ISP     &                30.00\% &              20.00\% &              30.00\% &            15.00\% &                 5.00\% \\
ASKED USER TO STOP          &                24.87\% &              26.42\% &              25.91\% &            17.10\% &                 5.70\% \\
SOLICITED COMMUNITY SUPPORT &                 5.36\% &              19.64\% &              33.93\% &            28.57\% &                12.50\% \\
CONSULTED LEGAL COUNSEL     &                12.50\% &              12.50\% &              37.50\% &            25.00\% &                12.50\% \\
REPORTED TO MAINTAINERS     &                10.53\% &              11.58\% &              32.63\% &            31.58\% &                13.68\% \\
CONTACTED LAW ENFORCEMENT   &                44.44\% &               0.00\% &              22.22\% &             0.00\% &                33.33\% \\
BLOCKED USER                &                 3.57\% &              11.90\% &              16.67\% &            33.33\% &                34.52\% \\
OTHER                       &                13.79\% &               0.00\% &              13.79\% &            34.48\% &                37.93\% \\
\bottomrule
\end{tabular}
}

    \begin{center}
    \adjustimage{max size={1\linewidth}{0.9\paperheight}}{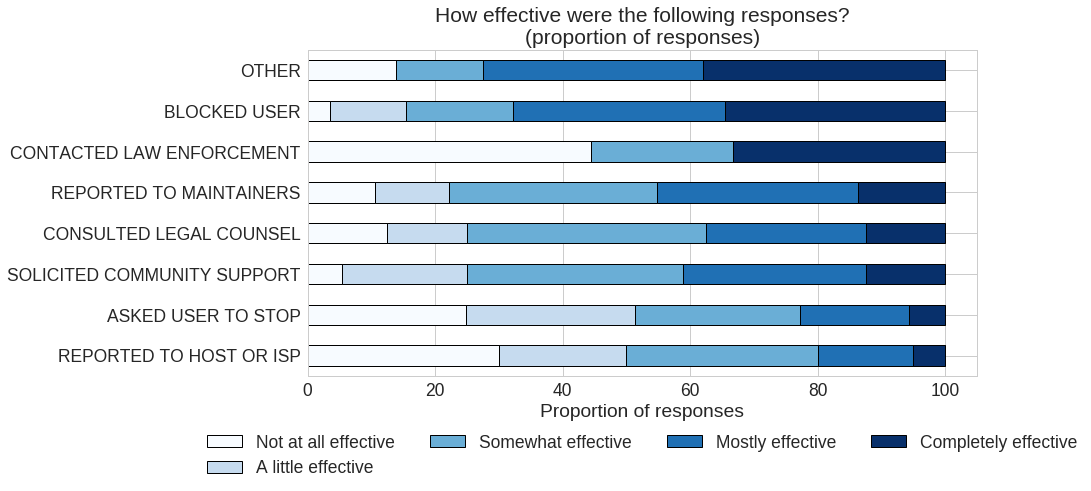}
    \end{center}
    { \hspace*{\fill} \\}
    
    \clearpage     \subsubsection{As a result of experiencing or witnessing harassment,
which, if any, of the following have you
done?}\label{as-a-result-of-experiencing-or-witnessing-harassment-which-if-any-of-the-following-have-you-done}

NEGATIVE.CONSEQUENCES.*

\texttt{\color{outcolor}Out[{\color{outcolor}193}]:}
    
    \centering{\begin{tabular}{lrr}
\toprule
{} &   Yes &    No \\
\midrule
NEGATIVE.CONSEQUENCES.STOPPED.CONTRIBUTING         &   390 &  1563 \\
NEGATIVE.CONSEQUENCES.PSEUDONYM                    &    50 &  1903 \\
NEGATIVE.CONSEQUENCES.WORK.IN.PRIVATE              &   166 &  1787 \\
NEGATIVE.CONSEQUENCES.CHANGE.USERNAME              &    48 &  1905 \\
NEGATIVE.CONSEQUENCES.CHANGE.ONLINE.PRESENCE       &    79 &  1874 \\
NEGATIVE.CONSEQUENCES.SUGGEST.COC                  &   116 &  1837 \\
NEGATIVE.CONSEQUENCES.PRIVATE.COMMUNITY.DISCUSSION &   301 &  1652 \\
NEGATIVE.CONSEQUENCES.PUBLIC.COMMUNITY.DISCUSSION  &   248 &  1705 \\
NEGATIVE.CONSEQUENCES.OFFLINE.CHANGES              &    85 &  1868 \\
NEGATIVE.CONSEQUENCES.OTHER                        &    90 &  1863 \\
NEGATIVE.CONSEQUENCES.NONE.OF.THE.ABOVE            &  1094 &   859 \\
\bottomrule
\end{tabular}
}

\texttt{\color{outcolor}Out[{\color{outcolor}194}]:}
    
    \centering{\begin{tabular}{lr}
\toprule
{} &  percent\_yes \\
\midrule
NEGATIVE.CONSEQUENCES.CHANGE.USERNAME              &        2.46\% \\
NEGATIVE.CONSEQUENCES.PSEUDONYM                    &        2.56\% \\
NEGATIVE.CONSEQUENCES.CHANGE.ONLINE.PRESENCE       &        4.05\% \\
NEGATIVE.CONSEQUENCES.OFFLINE.CHANGES              &        4.35\% \\
NEGATIVE.CONSEQUENCES.OTHER                        &        4.61\% \\
NEGATIVE.CONSEQUENCES.SUGGEST.COC                  &        5.94\% \\
NEGATIVE.CONSEQUENCES.WORK.IN.PRIVATE              &        8.50\% \\
NEGATIVE.CONSEQUENCES.PUBLIC.COMMUNITY.DISCUSSION  &       12.70\% \\
NEGATIVE.CONSEQUENCES.PRIVATE.COMMUNITY.DISCUSSION &       15.41\% \\
NEGATIVE.CONSEQUENCES.STOPPED.CONTRIBUTING         &       19.97\% \\
NEGATIVE.CONSEQUENCES.NONE.OF.THE.ABOVE            &       56.02\% \\
NEGATIVE.CONSEQUENCES.ANY.RESPONSE                 &      100.00\% \\
\bottomrule
\end{tabular}
}

    \begin{center}
    \adjustimage{max size={0.9\linewidth}{0.9\paperheight}}{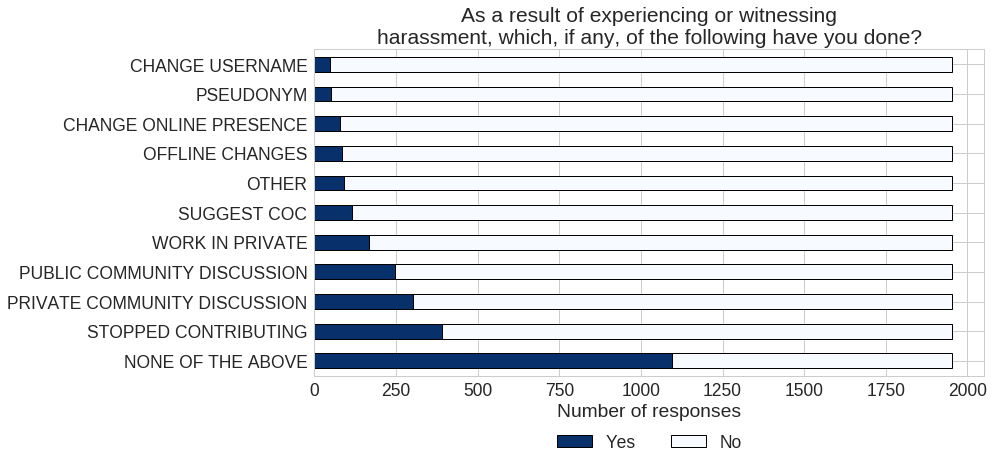}
    \end{center}
    { \hspace*{\fill} \\}

    
        \section{Bibliography}
\bibliographystyle{plainnat}
    
\bibliography{library.bib}

\begin{thebibliography}{11}
\providecommand{\natexlab}[1]{#1}
\providecommand{\url}[1]{\texttt{#1}}
\expandafter\ifx\csname urlstyle\endcsname\relax
  \providecommand{\doi}[1]{doi: #1}\else
  \providecommand{\doi}{doi: \begingroup \urlstyle{rm}\Url}\fi

\bibitem[Geiger(2017)]{geiger-report}
R.~Stuart Geiger.
\newblock Summary analysis of the 2017 github open source survey.
\newblock \emph{SocArXiv Preprints}, 2017.
\newblock \doi{10.17605/OSF.IO/ENRQ5}.
\newblock URL \url{https://osf.io/preprints/socarxiv/qps53}.

\bibitem[Hunter(2007)]{Matplotlib}
J.~D. Hunter.
\newblock Matplotlib: A 2d graphics environment.
\newblock \emph{Computing in Science Engineering}, 9\penalty0 (3):\penalty0
  90--95, May 2007.
\newblock ISSN 1521-9615.
\newblock \doi{10.1109/MCSE.2007.55}.
\newblock URL \url{http://ieeexplore.ieee.org/document/4160265/}.

\bibitem[Jones et~al.(2001)Jones, Oliphant, Peterson, et~al.]{scipy}
Eric Jones, Travis Oliphant, Pearu Peterson, et~al.
\newblock {SciPy}: Open source scientific tools for {Python}, 2001.
\newblock URL \url{http://www.scipy.org/}.

\bibitem[Kluyver et~al.(2016)Kluyver, Ragan-Kelley, P{\'e}rez, Granger,
  Bussonnier, Frederic, Kelley, Hamrick, Grout, Corlay, Ivanov, Avila, Abdalla,
  Willing, and development team]{jupyter}
Thomas Kluyver, Benjamin Ragan-Kelley, Fernando P{\'e}rez, Brian Granger,
  Matthias Bussonnier, Jonathan Frederic, Kyle Kelley, Jessica Hamrick, Jason
  Grout, Sylvain Corlay, Paul Ivanov, Dami{\'a}n Avila, Safia Abdalla, Carol
  Willing, and Jupyter development team.
\newblock Jupyter notebooks: a publishing format for reproducible computational
  workflows, 2016.
\newblock URL \url{https://eprints.soton.ac.uk/403913/}.

\bibitem[McKinney(2010)]{pandas}
Wes McKinney.
\newblock {Data Structures for Statistical Computing in Python}.
\newblock In St{\'{e}}fan van~der Walt and Jarrod Millman, editors,
  \emph{Proceedings of the 9th Python in Science Conference}, pages 51--56,
  2010.
\newblock URL
  \url{http://conference.scipy.org/proceedings/scipy2010/mckinney.html}.

\bibitem[P\'erez and Granger(2007)]{ipython}
Fernando P\'erez and Brian~E. Granger.
\newblock {IP}ython: a system for interactive scientific computing.
\newblock \emph{Computing in Science and Engineering}, 9\penalty0 (3):\penalty0
  21--29, May 2007.
\newblock ISSN 1521-9615.
\newblock \doi{10.1109/MCSE.2007.53}.
\newblock URL \url{http://ipython.org}.

\bibitem[van~der Walt et~al.(2011)van~der Walt, Colbert, and Varoquaux]{numpy}
S.~van~der Walt, S.~C. Colbert, and G.~Varoquaux.
\newblock The numpy array: A structure for efficient numerical computation.
\newblock \emph{Computing in Science Engineering}, 13\penalty0 (2):\penalty0
  22--30, March 2011.
\newblock ISSN 1521-9615.
\newblock \doi{10.1109/MCSE.2011.37}.
\newblock URL \url{https://arxiv.org/abs/1102.1523}.

\bibitem[van Rossum(1995)]{python}
Guido van Rossum.
\newblock Python library reference, 1995.
\newblock URL \url{https://ir.cwi.nl/pub/5009/05009D.pdf}.

\bibitem[Waskom et~al.(2014)Waskom, Botvinnik, Hobson, Cole, Halchenko, Hoyer,
  Miles, Augspurger, Yarkoni, Megies, Coelho, Wehner, cynddl, Ziegler,
  diego0020, Zaytsev, Hoppe, Seabold, Cloud, Koskinen, Meyer, Qalieh, and
  Allan]{seaborn}
Michael Waskom, Olga Botvinnik, Paul Hobson, John~B. Cole, Yaroslav Halchenko,
  Stephan Hoyer, Alistair Miles, Tom Augspurger, Tal Yarkoni, Tobias Megies,
  Luis~Pedro Coelho, Daniel Wehner, cynddl, Erik Ziegler, diego0020, Yury~V.
  Zaytsev, Travis Hoppe, Skipper Seabold, Phillip Cloud, Miikka Koskinen, Kyle
  Meyer, Adel Qalieh, and Dan Allan.
\newblock seaborn: v0.5.0 (november 2014), November 2014.
\newblock URL \url{https://doi.org/10.5281/zenodo.12710}.

\bibitem[Zlotnick et~al.(2017{\natexlab{a}})Zlotnick, Smith, Linksvayer,
  Filippova, and Initiative]{Github2017}
Frannie Zlotnick, Arfon Smith, Mike Linksvayer, Anna Filippova, and The
  Open~Source Initiative.
\newblock The open source survey.
\newblock \emph{GitHub repository}, 2017{\natexlab{a}}.
\newblock URL \url{https://github.com/github/open-source-survey}.

\bibitem[Zlotnick et~al.(2017{\natexlab{b}})Zlotnick, Smith, Linksvayer,
  Filippova, and Initiative]{Github2017-report}
Frannie Zlotnick, Arfon Smith, Mike Linksvayer, Anna Filippova, and The
  Open~Source Initiative.
\newblock The open source survey website.
\newblock \emph{GitHub, Inc.}, 2017{\natexlab{b}}.
\newblock URL \url{https://opensourcesurvey.org/2017/}.

\end{thebibliography}
    
    \end{document}